\documentclass[aps,prx,reprint,groupedaddress,twocolumn,showkeys,amsmath,amssymb,longbibliography,floatfix]{revtex4-2}

\usepackage{amssymb,amsmath,slashed,comment,amsthm}
\usepackage{hyperref}
\usepackage{color}
\usepackage{graphicx}
\usepackage{tabularx}
\usepackage{slashed}
\usepackage{tikz}
\usepackage{comment}
\usepackage{makecell}
\usepackage[normalem]{ulem}
\usepackage{transparent}
\usepackage{circledsteps}
\usepackage{subfigure}
\usepackage{array}
\def\l{\lambda}

\def\a{\alpha}

\def\p{\partial}
\def\m{\mathcal}
\def\D{\Delta}
\def\d{\delta}
\def\T{\Theta}
\def\s{\sigma}
\def\o{\omega}
\def\t{\theta}
\def\b{\boldsymbol}
\def\g{\omega}
\def\up{\uparrow}
\def\down{\downarrow}
\def\O{\Omega}

\newcommand{\vect}[1]{\boldsymbol{#1}}

\begin{document}

\title{Exotic phase transitions in spin ladders with discrete symmetries\\ that emulate spin-1/2 bosons in two dimensions
}

\author{Bo Han}
\author{David F. Mross} 
\affiliation{
Department of Condensed Matter Physics, Weizmann Institute of Science, Rehovot 7610001, Israel
            }

\date{\today}

\begin{abstract}
We introduce a spin ladder with discrete symmetries designed to emulate a two-dimensional spin-1/2 boson system at half-filling. Using global properties, such as the structure of topological defects, we establish a correspondence between the two systems and construct a dictionary of symmetries and operators. In particular, translation invariance leads to Lieb-Schultz-Mattis constraints for both systems, resulting in exotic deconfined quantum critical points. Subsequently, we study the spin ladder in detail. An exact duality transformation maps it onto a $\mathbb{Z}_2$ gauge theory of three partons, analogous to the U(1) gauge theory of chargons and spinons in two-dimensional spin-1/2 boson systems. With the mapping between spins and partons, we construct exactly solvable models for all pertinent symmetry-breaking phases and analyze their transitions. We further
make connections between our exact analysis and conventional parton gauge theories.

\end{abstract}
\maketitle

\section{Introduction}
\label{sec:Intro}

Characterizing phases of matter and their transitions is a cornerstone of condensed matter physics. Conventional phases, also known as `Landau phases', are fully specified by their (broken) symmetries, encoded in a local order parameter. Any symmetry-breaking pattern is associated with a specific set of topological defects (e.g., domain walls, vortices, or skyrmions) that depend on the space dimension, the microscopic symmetry group, and the extent to which it is broken in the ground state. Each individual defect connects regions where the order parameter takes different values, disrupting the uniform order. The proliferation of defects destroys the ordered state entirely and restores the symmetry, resulting in a transition to a different phase.

When transitions between different phases occur continuously, many observables follow universal scaling laws that are insensitive to the microscopic details. Landau's theory of phase transitions provides conditions under which such transitions can occur~\cite{Landau1937phasetransition}. In particular, it predicts that direct transitions between two phases that break distinct symmetries are generically first-order and require fine-tuning to be continuous. Alternatively, the two phases can be separated by an intermediate phase where both orders are present or absent.

Over the past few decades, however, it has been realized that Landau's theory is incomplete in characterizing phases and transitions. In particular, generic second-order transitions between phases that break distinct symmetries have been discovered~\cite{Senthil2004DQCP,Senthil2023dqcpreview}. Here, the topological defects in the order parameter on one side of the transitions carry quantum numbers of the symmetry that is broken on the other side. Consequently, restoring the first symmetry via defect proliferation immediately establishes the second order. The prototypical example of such a transition is between the N\'eel antiferromagnetic and valence-bond-solid (VBS) orders in two-dimensional spin-1/2 systems on square lattices~\cite{Senthil2004DQCP,Senthil2004prbdqcp,LevinSenthil2004dqcp,Sandvik2007dqcp}. 
In particular, the VBS defect carries spin-1/2, so that its proliferation breaks spin rotation symmetry, realizing a continuous transition into the antiferromagnet. Recently, similar Landau-forbidden transitions between magnetically ordered and VBS states were discovered in spin chains with two $\mathbb{Z}_2$ symmetries~\cite{JiangMotrunich2019DQCP,ZhangLevin2023dqcp1d,RobertsJiangMotrunich2019dqcp,Huang2019dqcp,Mudry2019qpt1d}. Analogous to the  N\'eel-VBS transition in 2D, the VBS defects in these spin chains also carry spin-1/2, and their condensation drives a Landau-forbidden phase transition.  For a recent review with additional examples, see Ref.~\onlinecite{Senthil2023dqcpreview}.

The similarity between these phase transitions, despite their different internal symmetries -- continuous SU(2) or U(1) in 2D vs. discrete $\mathbb{Z}_2 \times \mathbb{Z}_2$ in 1D -- is no coincidence. Ultimately, they relate to the fact that both systems exhibit symmetries that are incompatible with a unique symmetric gapped ground state without breaking translation invariance. Such constraints are known as Lieb-Schultz-Mattis (LSM) theorems, following original work by LSM and their generalizations~\cite{LieSchultzMattis1961LSM,Oshikawa2000LSM,Hastings2004LSM}. Recently, LSM theorems have been discussed in various contexts in one and two dimensions~\cite{Watanabe2015filling,Po2017magnet,Lu2017filling,Lu2024LSM,Yang2018lsm,MetlitskiThorngren2018anomaly,ChengSeiberg2023lsm,ChoHsiehRyu2017lsm,YaoHsiehOshikawa2019lsm,Seiberg2024lsm,YaoOshikawa2021LSM,Aksoy2024lsm,Cheng2019flsm,JianBiXu2018lsm}. These theorems imply that defects of one symmetry carry quantum numbers of the other. Consequently, destroying the order on one side of the critical point by defect proliferation necessarily induces a different order on the other side.

Close connections between $\mathbb{Z}_2$ symmetries in 1D and U(1) symmetries in 2D are well known in additional context. In particular, both are at the lower critical dimensions, and their defects are point-like objects, i.e., domain-walls in 1D and vortices in 2D. Consequently, both systems permit \textit{dual} descriptions, which treat defects as fundamental objects. In the 1D $\mathbb{Z}_2$ and 2D U(1) cases, these formulations are known as the Kramers-Wannier~\cite{KramersWannier1941duality} and  Peskin-Dasgupta-Halperin dualities~\cite{Peskin1978duality,DasguptaHalperin1981duality}, respectively. 

At phase transition points, a third formulation that treats both of the involved phases democratically is often useful. The composite of order parameter and defect is a fermion in both cases, i.e., a real (Majorana) fermion in 1D and a complex (Dirac) fermion in 2D. The fermions are coupled to dynamical $\mathbb{Z}_2$ and $U(1)$ gauge fields, respectively. This distinction is significant for scaling behavior, which is not modified by $\mathbb{Z}_2$ gauge fields but is strongly affected in the $U(1)$ case. Despite this difference, the 1D--2D correspondence extends to the gauge theory description. Specifically, the two fermion systems, 1D Majorana and 2D Dirac,  carry $\mathbb{Z}_2$ anomalies. A single Majorana fermion carries the \textit{chiral anomaly} characterized by the Arf-invariant~\cite{KarchTongTurner2019dualityweb2d}, while a single Dirac fermion carries the well-known \textit{parity anomaly} characterized by the Chern-Simons invariant~\cite{NiemiSemenoff1983parityanomlay,Redlich1984parity,Witten2016RMPfermion,Witten2016parityanomaly}. These anomalies become manifest in the partition functions in the limit of infinite fermion masses. Flipping the sign of the mass term results in an additional phase factor in their partition functions, determined by the respective topological invariants~\cite{KarchTongTurner2019dualityweb2d,CordovaFreedLamSeiberg2020anomaly1,Seiberg2016duality,KarchTong2016duality}, i.e., 
\begin{equation}
\begin{split}
  & \text{1D Majorana fermion }:~ \frac{Z[-m,\rho]}{Z[m,\rho]} = e^{i \pi \text{Arf}(\rho)}, \\
   & \text{2D Dirac fermion }:\quad\ \ ~ \frac{Z[-m,A]}{Z[m,A]} = e^{i \text{CS}(A)},  
\end{split}
\end{equation}
where $\rho$ denotes the $\mathbb{Z}_2$ background spin structure and $A$ is the U(1) background electromagnetic field.

In this work, we explore additional unconventional phase transitions and further connections between discrete and continuous symmetries in one and two dimensions. Specifically, we consider 2D systems of spin-1/2 bosons at half-filling with two $U(1)$ symmetries, spin and charge. We construct the corresponding 1D systems based on general considerations such as LSM theorems and the structure of topological defects and study them in detail. Introducing an exact reformulation of the microscopic spin model in terms of parton variables and gauge fields, we construct lattice models for various phases and transitions. Additionally, we compute topological responses and analyze phase transitions, in particular, how they are `conventionalized' when translation symmetry is broken explicitly.

The paper is organized as follows. In Section~\ref{sec:symmetries}, we introduce a spin ladder and its $\mathbb{Z}_2$ symmetries designed to emulate 2D spinful bosons at half-filling. We also formalize how to diagnose phases via response actions of background gauge fields. In Section~\ref{sec:simplephase}, we discuss several simple phases that can be understood from local spin representations and relate them to their 2D analogs. In Section~\ref{sec:parton} we introduce partons and dualities first at the level of field theory and then at the lattice.
In Section~\ref{sec:partonphase}, we utilize this parton representation to construct lattice Hamiltonians of various phases and describe their ground states and response properties. Section~\ref{sec:partontranslation} constructs translation invariant models that realize the phases where translation symmetry is broken spontaneously. Phase transitions are analyzed in Section~\ref{sec:transition}. We conclude in Section~\ref{sec:conclusion} with a brief summary of results and a discussion of future research directions they suggest. Several appendices contain additional technical details of our analysis and derivations.

\section{Symmetries and background gauge fields}
\label{sec:symmetries}
We want to construct a one-dimensional system that emulates phases and phase transitions of spin-1/2 bosons $b_{\uparrow,\downarrow}$ in two dimensions~\cite{Boninsegni2001HCB,Cole2012socBH}. The boson number of each spin is separately conserved, i.e., each is associated with a U(1) symmetry. As discussed above, the one-dimensional analog will not capture universal properties such as exponents quantitatively, much less microscopic aspects. Still, it is helpful to frame the discussion based on a specific boson Hamiltonian
\begin{equation}
\begin{split}
    H &= - t\sum_{\langle\b{r}\b{r}'\rangle,\l}  b^\dagger_{\b{r}\l} b_{\b{r}'\l}  - V\sum_{\b{r},\l}  n_{\b{r} \l} (n_{\b{r} \l} -1) \\
    & \quad + U\sum_{\b{r}} n_{\b{r} \up} n_{\b{r}\down}+ \sum_{\b{r} \neq \b{r}', \l} U'_{\b{r} \b{r}'} n_{\b{r}\l} n_{\b{r}'\l} .
    \end{split} \label{eqn:BoseHubbard}
\end{equation}

The first line represents a standard Bose-Hubbard model for each boson species, $\lambda=\up, \down$while the second line parameterizes interactions between opposite spins ($U$ term) and different sites ($U'$ term). 

When the single-species Hubbard interaction dominates ($V \to \infty$), each species is a hard-core boson, with the occupancies restricted to $n_\l = 0, 1$ are the only allowed occupancies. When additionally $U \to \infty$, the local Hilbert space is reduced to two states,  $|\!\!\uparrow_b\rangle\equiv|1_\uparrow, 0_\downarrow \rangle$ and $|\!\!\downarrow_b\rangle \equiv|0_\uparrow , 1_\downarrow \rangle$ in the occupation number basis $|n_\uparrow,n_\downarrow\rangle$. Consequently, this limit results in an effective spin-1/2 system. At intermediate $U,V$, a chemical potential is necessary to maintain half-filling $\sum_\l \langle n_\l\rangle =1$.

In the one-dimensional context, continuous U(1) symmetries are replaced with discrete $\mathbb{Z}_2$ symmetries. Specifically, we encode the internal quantum numbers $\uparrow,\downarrow$ of the bosons as the upper and lower legs of a spin ladder. The two `$\mathbb{Z}_2$ Charge' symmetries are generated by
\begin{equation}
\begin{split}
    &~ g^\uparrow_x \equiv \prod\nolimits_r \s^x_{\uparrow,r}~, \qquad\qquad{\text{(Charge-$\uparrow$)}} \\
    &~ g^\downarrow_x \equiv \prod\nolimits_r \s^x_{\downarrow,r}~,\qquad\qquad{\text{(Charge-$\downarrow$)}}
    \end{split}
   \label{eqn:Z2sa}
\end{equation}
i.e., the boson densities $n_\l$ correspond to the spin component $\sigma^x_\l$. The half-filling property of the 2D bosons does not require any symmetry and has no direct analog in a $\mathbb{Z}_2$ system. However, \textit{hard-core} bosons may exhibit a particle-hole symmetry ${\cal C} : n_{\uparrow,\downarrow} \rightarrow 1- n_{\uparrow,\downarrow}$ that requires half-filling. In particular, this symmetry interchanges the states $|\!\!\uparrow_b\rangle,|\!\!\downarrow_b\rangle$, realizing a spin flip.
The analogous action on the $\sigma^x$ eigenstates is performed by $\sigma^z_\uparrow \sigma^z_{\downarrow}$. Thus, 
we identify particle-hole symmetry $\m{C}$ on the spin ladder with
\begin{equation}
    g_z \equiv \prod\nolimits_r (\s^z_{\uparrow,r} \s^z_{\downarrow,r})~ \qquad\qquad{\text{(Spin)}}
   \label{eqn:Z2sb}.
\end{equation}

Hard-core bosons at half-filling are subject to LSM constraints~\cite{LieSchultzMattis1961LSM,Oshikawa2000LSM,Hastings2004LSM}, which follow from the anticommutation between particle-hole symmetry and the U(1) particle number conservation within each unit cell. Consequently, they cannot form a trivial gapped ground state without breaking symmetries.  
Similarly, the symmetries $g_z$ and $g^{\uparrow,\downarrow}_x$ anticommute within each unit cell of the spin ladder; see also Appendix~\ref{appendix:LSM}. Consequently, a symmetric gapped ground state is incompatible with the lattice translation symmetry 
\begin{align}
T_x: &\b{\s}_{\l,r} \mapsto \b{\s}_{\l,r+1}.
\end{align}
Specifically, a gapped and translation symmetric ground state must break Spin symmetry $(g_z)$, or \textit{both} Charge symmetries $(g^{\up,\down}_x)$. In addition to translations and $g_{x,z}$, we also include a reflection symmetry ${\cal R}$ that interchanges the legs of the ladder, and corresponds to an internal $\pi$ rotation of the boson spin. We summarize the correspondence between the symmetries and operators of the 2D bosons and the spin ladder in Table~\ref{tab:dict}. 

\begin{table}
    \centering
     \caption{Correspondence between the symmetries and operators spin-1/2 bosons and spin ladder. $b_\l$ are the annihilation operators of spin  $\l = \uparrow,\downarrow$ bosons and $n_\l = b^\dagger_\l b_\l$ their number operators. Notice that the $\mathbb{Z}_2$ nature makes the creation and destruction of the $\mathbb{Z}_2$ charge indistinguishable, namely, $(\s^z)^\dagger = \s^z$. $\m{R}$ acts as a $\pi$ rotation symmetry of the boson spin and corresponds to reflection symmetry that interchanges the legs of the spin ladder.}
    \begin{tabular}{c|cc}
    \hline \hline
        & 2D spinful boson & 1D spin-1/2 ladder \\ 
    \hline    
       & U(1)$_\uparrow$ & $\mathbb{Z}_{2 \uparrow}^x$ \\ 
       Symmetries &  U(1)$_\downarrow$ & $\mathbb{Z}_{2 \downarrow}^x$ \\ 
        & half-filling & $\mathbb{Z}_2^z\ $ \\ 
        & $\m{R}$: $b_\up \leftrightarrow b_\down$ & $\m{R}$: $\s_\uparrow \leftrightarrow \s_\downarrow$ \\
        \hline
          &  $n_\l$ & $\sigma^x_\l$ \\ 
        Operators  & $n_\up + n_\down$ & $\s^x_\up \s^x_\down$ \\
       &  $b_\l$, $b^{\dagger}_\l$ & $\s^z_\l$ \\ 
       &  $b^\dagger_\uparrow b_\downarrow$ & $\s^z_\uparrow \s^z_\downarrow$ \\ 
    \hline \hline
    \end{tabular}
    \label{tab:dict}
\end{table}

\subsection{Background gauge fields}
\label{sec:couplingbackgroundfield}
A convenient way to identify gapped phases is by their response to background gauge fields associated with their symmetries. Schematically, we couple a theory of a matter field $\phi$ with conserved current $\b{j}_\phi$ to a background field $\vect{A}$ by the action
\begin{align}
   S[\phi,\b{A}] &= S[\phi] + \int \b{j}_\phi \cdot \b{A}. 
\end{align}
Since $\phi$ is in a gapped phase, we integrate it out to obtain the effective action of the background field 
\begin{align}
    e^{i S_{\text{eff}}[\b{A}]} &= \int \m{D}\phi \ e^{i S[\phi,\b{A}]}.
\end{align}
The response action $S_{\text{eff}}[\b{A}]$ effectively diagnoses the phase of the system. 

Specifically, an effective action proportional to a Maxwell term,
i.e., $S_{\text{eff}}[\b{A}] \propto \frac{1}{m^2_\phi} \int (\nabla \times \b{A})^2$, indicates that $\phi$ forms an insulator with gap $m_\phi$. In the $m_\phi \to \infty$ limit, the induced action vanishes, and the system does not respond to the background field. 

By contrast, when $\phi$ forms a condensate, the effective action contains a mass term proportional to the superfluid stiffness $\rho_s$ for the transverse component $\b{A}_\perp$ of the background field, i.e., $S_{\text{eff}}[\b{A}] \sim \rho_s\int  \b{A}^2_\perp$. As a result, the gauge flux of $\b{A}$ is expelled, i.e., the system exhibits a Meissner effect.

A topologically nontrivial state in two dimensions can lead to a Chern-Simons response $S_{\text{eff}}[\b{A}] \propto \int A_I d A_J$, where $A_I$ are different flavors of U(1) background fields. 
The topological response implies that inserting a flux quantum of the gauge field $A_J$ accumulates a quantized charge of the gauge field $A_I$. 

One-dimensional systems only permit global fluxes due to the absence of locally closed loops. Consequently, their flux configurations are fully specified by boundary conditions. Determining the background response is thus equivalent to identifying the flux sector (boundary condition) of the system's ground state(s).

To apply this approach, we introduce background $\mathbb{Z}_2$ gauge fields $\O_a$ with $a =$ C$\up$,~C$\down$,~S for Charge and Spin symmetries~[Eq.~(\ref{eqn:Z2sa}) and (\ref{eqn:Z2sb})], which couple to the specific terms in the spin ladder as
\begin{equation}
\begin{split}
\s^z_{\up,r} \s^z_{\up,r+1} &\rightarrow \s^z_{\up,r} \O^z_{\text{C} \up, r+1/2} \s^z_{\up,r+1}, \\
\s^z_{\down,r} \s^z_{\down,r+1} &\rightarrow \s^z_{\down,r} \O^z_{\text{C} \down, r+1/2} \s^z_{\down,r+1}, \\
\s^x_{\l,r} \s^x_{\l,r+1}  &\rightarrow \s^x_{\l,r} \O^z_{\text{S}, r+1/2} \s^x_{\l,r+1}, \quad \l = \up, \down.
\end{split} \label{eqn:minimalcoupling}
\end{equation}
Integrating out the local spin variables leads to effective actions of the background gauge fields. An insulator is defined by an effective action that imposes no constraints on the background gauge flux $\Phi_a = \prod \O_a $, analogous to the Maxwell response. The effective action of $\mathbb{Z}_2$ condensates imposes constraints on the flux, indicating Meissner effects for these $\mathbb{Z}_2$ gauge fields, i.e., flux expulsion analogous to U(1) condensates.  Finally, the effective action of symmetry-protected topological (SPT) states ties the background gauge flux to a conserved charge $\Phi_a  = g_b$, mirroring the mutual Chern-Simons terms for U(1) gauge fields.

Previous applications of background $\mathbb{Z}_2$ gauge fields include supersymmetric systems coupled to a fermion-parity gauge field to diagnose spontaneous supersymmetry breaking~\cite{Witten1982wittenindex}; Majorana chains coupled to a background spin structure to detect the topological response~\cite{Shiozaki2017fspt}; finite-temperature topological phases coupled to background gauge fields for classification purposes~\cite{HuangDiehl2024finiteT}.  

\section{Simple phases}
\label{sec:simplephase}
We begin the analysis of spin ladders with Charge and Spin symmetries defined in Eq.~(\ref{eqn:Z2sa}) and (\ref{eqn:Z2sb}) by discussing several simple phases. 

\subsection{Decoupled limit}
\label{subsec:decoupled}
\begin{figure}[h]
    \centering
    \includegraphics[width=0.4\textwidth]{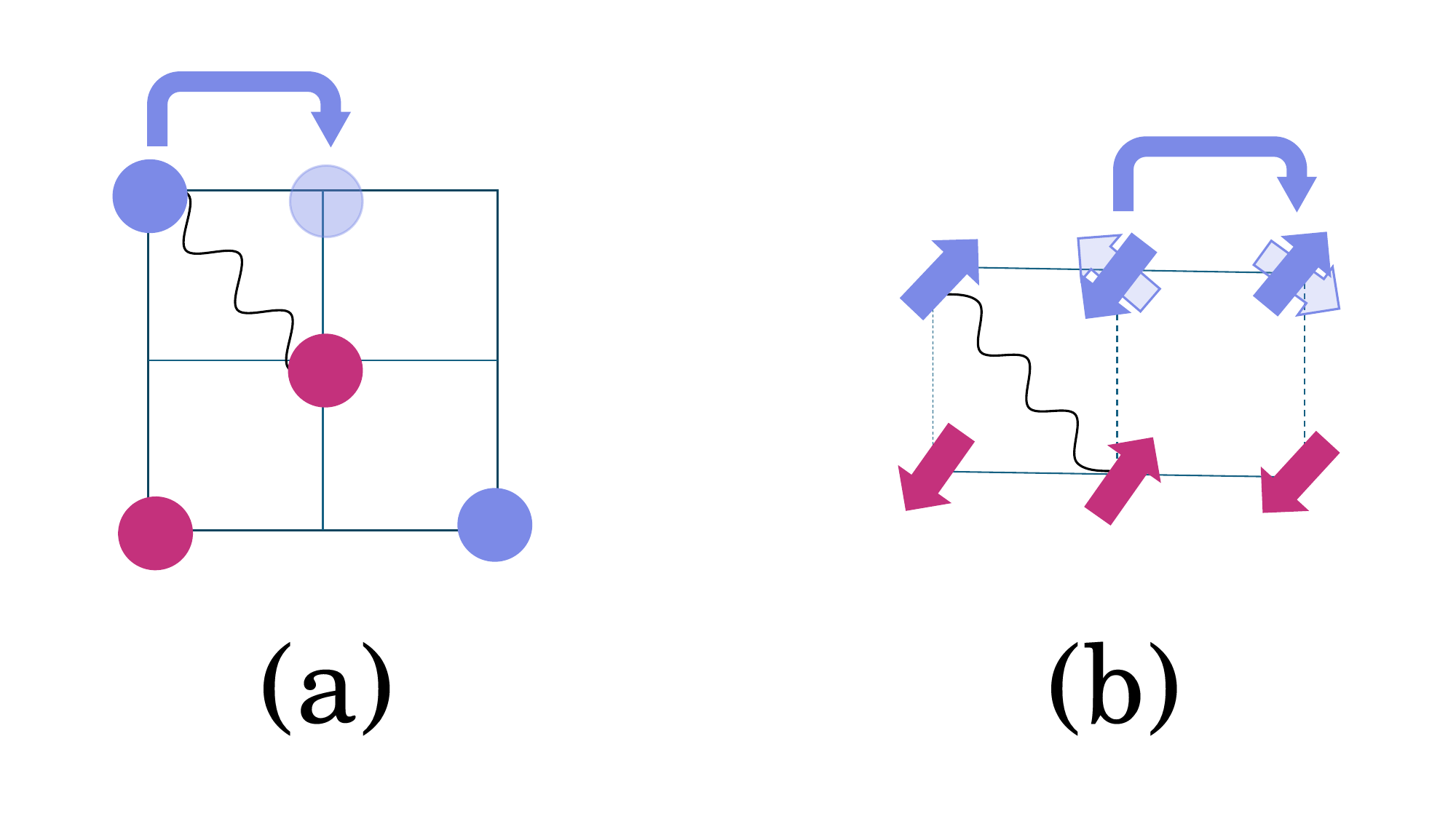}
    \caption{(a)~Spinful bosons with $U(1)_\uparrow \times U(1)_\downarrow$ symmetries at half-filling on the square lattice. Blue and pink circles represent spin-up and spin-down bosons, respectively. The blue arrow indicates bosons hopping between neighboring sites and the black wavy line represents interactions between bosons. (b)~Spin-1/2 ladder with $\mathbb{Z}_{2\uparrow} \times \mathbb{Z}_{2\downarrow}$ symmetries. The short wide arrows on each site indicate lattice spins. The dark and light blue shades show their state before and after a spin flip, which is analogous to boson hopping (a). The black wavy line represents interchain interactions.}
    \label{fig:1d2d}
\end{figure}
Without interchain couplings, the Spin symmetry of Eq.~(\ref{eqn:Z2sb}) factorizes into two independent symmetries---one for each leg. Within a unit cell of the  leg $\lambda$, the symmetries $\mathbb{Z}^x_{2,\lambda}$ and $\mathbb{Z}^z_2$ anti-commute, implying an LSM constraint. Consequently, a symmetric gapped ground state is impossible. Instead, there are ordered phases (characterized by spontaneous breaking of onsite symmetries) and VBS phases (characterized by spontaneous breaking of lattice symmetries).  These phases are separated by continuous phase transitions governed by deconfined quantum critical points, as explored in Ref.~\onlinecite{JiangMotrunich2019DQCP} in a single Ising chain. 

\subsection{Valence bond solid phases}
\label{subsec:VBS}
Valence bond solids preserve all internal symmetries but spontaneously break translation symmetry~\cite{ReadSachdev1989vbs,AKLT1987}. In the simplest incarnations of these phases, pairs of neighboring spins form a symmetric quantum state. There are two VBS states for each leg, which are related by lattice translations. Phase boundaries between the two VBS configurations host unpaired spins. Consequently, the decoupled spin ladder has four distinct VBS states, which are shown in Fig.~\ref{fig:VBSSPT} below.
\begin{figure}[h]
\centering
\includegraphics[scale = 0.5]{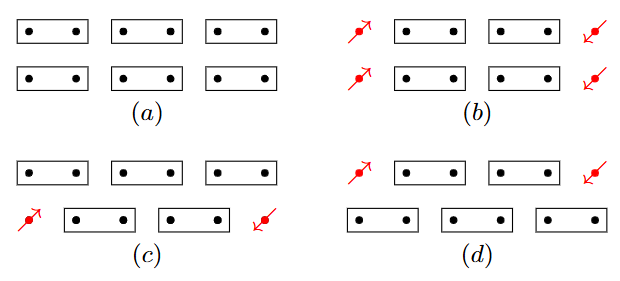}
\caption{Distinct VBS states for the spin ladder. Without coupling between the two legs, four VBS states are degenerate. Spatial boundaries between different VBS states host local spin degrees of freedom (red arrows) that cannot be gapped without breaking the internal symmetries of the ladder.} 
\label{fig:VBSSPT}
\end{figure}
\begin{table}
\centering
\caption{Partition functions $Z[g,h^*]$ for the four distinct topological phases with $h^*=(1,-1,-1)$ and different twisted boundary conditions characterized by $g$ along the temporal direction; see Appendix~\ref{app:VBS} for details.}
\begin{tabular}{ c|cccc } 
\hline \hline
$g$ & $(a)$ & $(b)$ & $(c)$ & $(d)$ \\
\hline
$g^\up_x$ &+1 & $-1$ & $-1$ & +1 \\
$g^\down_x$ &+1 & $-1$ & +1 & $-1$ \\
$g_z$ & +1 & $-1$ & $-1$ & +1 \\
\hline \hline
\end{tabular}
\label{tab:pf}
\end{table}

Interfaces between any two of these states cannot be gapped by interactions that preserve the onsite $\mathbb{Z}_2$ symmetries in Eq.~(\ref{eqn:Z2sa}) and (\ref{eqn:Z2sb}). The single unpaired spin at an interface between (a,c) or (a,d) in Fig.~\ref{fig:VBSSPT} cannot be gapped out by Charge and Spin symmetry-preserving operators. Interfaces with two dangling spins, i.e., those between (a,b) or (c,d) in Fig.~\ref{fig:VBSSPT} permit a single symmetry-allowed interaction,  $\s^x_\up \s^x_\down$, which leaves a two-fold degeneracy. 
Consequently, the four states in Fig.~\ref{fig:VBSSPT} cannot be smoothly deformed into one another without gap closure. 
This sharp distinction does not require translation symmetry. From the viewpoint of topological phases, the four states are
SPTs protected by the $\mathbb{Z}_2$ symmetries in Eq.~(\ref{eqn:Z2sa}) and (\ref{eqn:Z2sb})~\cite{GuWen2009spt,Senthil2015sptreview}. To demonstrate this property explicitly, we evaluate the partition function 
\begin{align}
    Z[g;h] &\equiv \text{Tr}_{\m{H}_h} \left( g e^{-\beta H} \right),
\end{align}
with boundary conditions twisted by $g = (g^\uparrow_x,g^\downarrow_x,g_z)$ and $h = (h^\uparrow_x,h^\downarrow_x,h_z)$ along the temporal and spatial cycles of the spacetime torus, respectively~\cite{DijkgraafVafaVerlindeVerlinde1989orbifold}; see Table~\ref{tab:pf}. The distinct values for three choices of the boundary conditions uniquely identify each of the four configurations and demonstrate that they represent different SPTs.

\subsection{$\mathbb{Z}_2$ charge insulators}
\label{subsec:z2ci}

The bosonic analogs of the VBS states (a) and (b) in Fig.~\ref{fig:VBSSPT} are Mott insulators where fluctuations of the U(1) charge are suppressed. We refer to the corresponding phases with frozen $\mathbb{Z}_2$ charge $g^\uparrow_x g^\downarrow_x$ as `$\mathbb{Z}_2$ Charge insulators'.

A simple interaction that results in an insulator where the $\mathbb{Z}_2$ charge is frozen on each rung of the ladder is ~\footnote{Without loss of generality, all the coupling constants in this paper are taken to be positive.} 
\begin{align}
    H_U &=  -U \sum_r \s^x_{\uparrow,r} \s^x_{\downarrow,r}~. \label{eqn:z2ci}
\end{align} 
 It is closely analogous to the Hubbard interaction [$U$-term in Eq.~(\ref{eqn:BoseHubbard})], which fixes the particle number on each site in the $U~\rightarrow~\infty$ limit. Similarly, $H_{U\rightarrow \infty}$ projects onto a reduced Hilbert space with two states per site, 
which we represent by spin-1/2 operators
\begin{equation}
    \begin{split}
 S^x_r&=    \s^x_{\uparrow,r}=\s^x_{\downarrow,r}~,\\
 S^y_r&= \s^z_{\uparrow,r} \s^y_{\downarrow,r}=\s^y_{\uparrow,r} \s^z_{\downarrow,r}~, \\
 S^z_r&= \s^z_{\uparrow,r} \s^z_{\downarrow,r} =- \s^y_{\uparrow,r} \s^y_{\downarrow,r} ~.
    \end{split}
    \label{curly}
\end{equation}
The symmetries in Eqs.~\eqref{eqn:Z2sa} and \eqref{eqn:Z2sb}, projected into the reduced Hilbert space, are
\begin{align}
    g^{\uparrow(\downarrow)}_x |_{U \to \infty} &= \prod_r S^x_r~, \qquad g_z |_{U \to \infty} = \prod_r S^z_r~. \label{eqn:Z2seff}
\end{align}
The system represented by $S_r$ in the projected Hilbert space is the same as either leg of the spin ladder in the decoupled limit discussed in Section~\ref{subsec:decoupled}. Notice that the effective Ising chain with the $\mathbb{Z}_2$ symmetries in Eq.~(\ref{eqn:Z2seff}) and translation symmetry is identical to the one discussed in Ref.~\cite{JiangMotrunich2019DQCP}.   

While $\mathbb{Z}_2$ charge fluctuations are frozen, the $\mathbb{Z}_2$ Spin, Eq.~\eqref{eqn:Z2sb}, remains unspecified and can realize distinct states. 
In particular, either or both of the $\mathbb{Z}_2$ symmetries in Eq.~(\ref{eqn:Z2seff}) and the lattice translation symmetry can be broken spontaneously. 
When neither of the $\mathbb{Z}_2$ symmetries is broken, a nontrivial SPT state protected by $\mathbb{Z}_2 \times \mathbb{Z}_2$ symmetry is possible, but requires the breaking of translation symmetry due to the LSM constraint. The breaking of a single $\mathbb{Z}_2$ symmetry in Eq.~(\ref{eqn:Z2seff}) leads to a $g_z$-breaking phase ($x$FM) or a $g_x$-breaking phase ($z$FM). The breaking of both 
symmetries while preserving their product leads to the fourth symmetry-breaking phase ($y$FM).

\subsection{$\mathbb{Z}_2$ spin condensates}
\label{subsec:z2sc}
A second set of phases, overlapping with $\mathbb{Z}_2$ charge insulators, are $\mathbb{Z}_2$ spin condensates. They are defined by broken $\mathbb{Z}_{2 \uparrow}^x$ and $\mathbb{Z}_{2 \downarrow}^x$ symmetries, i.e., by the order parameter $\langle \s^z_\uparrow \s^z_\downarrow \rangle \neq 0$. The analogous phases of bosons exhibit easy-plane magnetic order $\langle b^\dagger_\uparrow b_\downarrow\rangle \neq 0$. An interaction that induces this order on the spin ladder is
\begin{align}
    H_{J} &= -J \sum_{\langle rr' \rangle} \s^z_{\uparrow,r} \s^z_{\downarrow,r} \s^z_{\uparrow,r'} \s^z_{\downarrow,r'},
\end{align} 
with the $U(1)$ analogue 
\begin{align}
    H_{J} &= -J \sum_{\langle \vect{r}\vect{r}' \rangle} b^\dagger_{\uparrow,\vect{r}} b_{\downarrow,\vect{r}} b^\dagger_{\downarrow,\vect{r}'}b_{\uparrow,\vect{r}'}  ~.
\end{align}
In the $\mathbb{Z}_2$ case, the ground states of $H_J$ decompose into two sectors labeled by the sign of the order parameter $P \equiv \s^z_\uparrow \s^z_\downarrow  = \pm 1$. For $J\rightarrow \infty$, each sector can be described by a spin-1/2 system represented by
\begin{align}
   \tilde{S}^z_r &= \s^z_{\uparrow,r} = P \s^z_{\downarrow,r}~, \ \qquad \tilde{S}^x_r = \s^x_{\uparrow,r} \s^x_{\downarrow,r}. \label{eqn:dictSS}
\end{align}
Since $g^\uparrow_x$ and $g^\downarrow_x$ flip the sign of $P$, they cannot be represented by $\tilde S$. The other symmetries act as
\begin{align}
    g_z\big|_{J \to \infty} &= \prod_r \tilde{S}^z_r~, \ \qquad g^\uparrow_x g^\downarrow_x \big|_{J \to \infty} = \prod_r \tilde{S}^x_r.\label{eqn.effS.syms}
\end{align}

The effective description is again equivalent to either leg of the original spin ladder in the decoupled limit. Nevertheless, the $\tilde{S}$ representation describes different phases and transitions since $g_x^\uparrow$ and $g_x^\downarrow$ of Eq.~(\ref{eqn:Z2sa}) are broken in all $\mathbb{Z}_2$ condensates, but remain possible symmetries of $\mathbb{Z}_2$ charge insulators.

\subsection{$\mathbb{Z}_2$ charge insulators with $\mathbb{Z}_2$ spin order}
To anticipate the nature of phase transitions, it is useful to study a phase accessible from the effective descriptions of Sections~\ref{subsec:z2ci} or  \ref{subsec:z2sc}. This phase is a $\mathbb{Z}_2$ charge insulator with $\mathbb{Z}_2$ spin order, which can be realized by
\begin{align}
    H &= H_U + H_J. \label{eqn:chargespin}
\end{align}
All terms in $H_U$ and $H_J$ commute and can be readily diagonalized in any order. Diagonalizing $H_U$ first, we project onto the Hilbert space represented by the operators $S$ of Eq.~(\ref{curly}). Diagonalizing $H_J$ first, we instead project onto the  Hilbert space represented by $\tilde S$  of Eq.~(\ref{eqn:dictSS}). We thus find
\begin{equation}\begin{split}&H_J|_{U \to \infty} = -J \sum_r S^z_r S^z_{r+1}\\&H_U|_{J \to \infty} = -U \sum_r \tilde{S}^x_r\end{split}\end{equation} The equivalent descriptions $H_J|_{U \to \infty}$ and $H_U|_{J \to \infty}$ are strongly reminiscent of dual formulations of transverse field Ising chains. This observation suggests that dual variables can provide a unified description of phases and transitions.

\section{Parton representations and global phases}
\label{sec:parton}

Sections~\ref{subsec:z2ci} and \ref{subsec:z2sc} analyzed Hamiltonians in which the $\mathbb{Z}_2$ charge and spin degrees of freedom decouple. A versatile framework for exploring such phases and their transitions is given by parton representations. We decompose the microscopic spins $\sigma$ as~\cite{SenthilFisher2000fractionization}
\begin{align}
\sigma^z_{\lambda,r} = \tau^z_{c,r}\tau^z_{n,\lambda,r}\qquad (\lambda=\uparrow,\downarrow)\label{eqn.parton}
\end{align}
into a charge parton (chargon) $\tau^z_{c}$ and spin parton (spinon) $\tau^z_{n,\lambda}$. This decomposition mirrors the common `slave-boson' representation of electrons $c_\l = b f_\l$~\cite{Fradkinbook,Sachdev2023,LeeNagaosaWen2006rmp}. The $\mathbb{Z}_2$ partons are subject to the local constraint 
\begin{align}
\sigma^x_{\uparrow,r} \sigma^x_{\downarrow,r}  = \tau^x_{c,r} = \tau^x_{n,\uparrow,r} \tau^x_{n,\downarrow,r}~.  
\end{align}
This constraint implements local gauge transformations under which $\tau_{a,r}^z \rightarrow -\tau^z_{a,r}$ with $a=c,n\uparrow,n\downarrow$.

The simplest kinetic Hamiltonian for the $\tau$ variables takes the form
\begin{align}
H_\tau=& -t_c \sum_{r} \tau^z_{c,r}\omega^z_{r+1/2}\tau^z_{c,r+1} \label{eqn.htau} \nonumber \\
&  -t_n \sum_{r,\lambda} \tau^z_{n,\lambda,r}\omega^z_{r+1/2}\tau^z_{n,\lambda,r+1}~,
\end{align}
where $\omega^z_{r+1/2}$ is a $\mathbb{Z}_2$ gauge field obeying the Gauss-law
\begin{align}
\omega^x_{r-1/2}\omega^x_{r+1/2}=\tau^x_{c,r}\tau^x_{n,\uparrow,r} \tau^x_{n,\downarrow,r}~.
\end{align}
To interpret the flux $\Phi_\omega \equiv \prod_r \omega^z_{r+1/2}$ of the emergent gauge field, we observe that
\begin{align}
g_z &=\Phi_\omega \times \prod_{r\text{ even}} (\tau^z_{n,\uparrow,r} \omega^z_{r+1/2} \tau^z_{n,\uparrow,r+1}) \nonumber \\
& \quad \times \prod_{r \text{ even}} (\tau^z_{n,\downarrow,r-1} \omega^z_{r-1/2} \tau^z_{n,\downarrow,r})~,
\end{align}
which follows from inserting Eq.~\eqref{eqn.parton} into Eq.~\eqref{eqn:Z2sb}. Each factor in the product on the right-hand side is independently gauge invariant and appears as a term in $H_\tau$. A nonzero expectation value of this product identifies the $\mathbb{Z}_2$ flux with the generator $g_z$ of Spin symmetry.

The product of the two $g_x$ symmetries enforces the separate conservation of spin and charge partons, i.e.
\begin{align}
        g^\uparrow_x g^\downarrow_x&= \prod_r \tau^x_{c,r} =  \prod_r (\tau^x_{n,\uparrow,r} \tau^x_{n,\downarrow,r} ). \label{eqn:gxconventional}
\end{align}
More specifically, transmutations between chargons and spinons via terms such as $\tau^z_c \tau^z_{n,\lambda}$ break total $\mathbb{Z}_2$ charge symmetry $g^\uparrow_x g^\downarrow_x$; see also Eq.~(\ref{eqn.parton}). The individual $\mathbb{Z}_2$ charge symmetries cannot be expressed using $\tau$ variables alone; they are given by
\begin{align}
g_x^{\lambda} = \prod_r \tau^x_{n,\lambda,r} \times \prod \omega^x_{r+1/2}~,
\end{align}
and enforce separate conservation of $\tau_{n,\uparrow}$ and $\tau_{n,\downarrow}$. We note that only two of the three conserved $\tau$-parton numbers (parities) are independent, consistent with the number of Charge symmetries.

On the mean-field level, the $\tau_c$ exhibits a $\mathbb {Z}_2$ symmetry, and the $\tau_{n,\l}$ exhibit a separate $\mathbb{Z}_2 \times \mathbb{Z}_2$ symmetry. We refer to phases where the chargon parity is preserved in the ground state as chargon insulators and those where it is spontaneously broken as chargon condensates. For the spinons, the ground state may break both, one, or neither of the $\mathbb{Z}_2$ symmetries. In the last case, the $\mathbb{Z}_2 \times \mathbb{Z}_2$ symmetry allows for a nontrivial SPT in the phase of the well-known cluster state~\cite{ChenGuLiuWen2013spt,SonAmicoVedral2012spt1d}. The eight combinations of chargon and spinon states are listed in Table~\ref{tab:phases}. Among the phases listed in Table~\ref{tab:phases}, three preserve all onsite symmetries and must, therefore, break translation symmetry due to LSM constraints. 

\begin{table}[h]
\centering
\caption{Phases and broken symmetries for different parton states. For the $y$FM phase, the three listed symmetries are broken individually, but any product of two of them is preserved. Double Condensate refers to the state where both spinons condense, while Condensate denotes the state where one spinon forms an insulator and the other consenses. 
} 
\begin{tabular}{ cccc } 
\hline\hline
Chargon & Spinons & Broken Symmetries & Phases   \\
\hline
Insulator & \makecell{Insulator \\ SPT State \\ Double Condensate \\  Condensate} &  \makecell{$g_z$ \\ $g^\uparrow_x, g^{\downarrow}_x, g_z$ \\ $g^\uparrow_x, g^\down_x$  \\ $T_x$ 
} & \makecell{$x$FM \\$y$FM \\ $z$FM \\ VBS} \\
\hline
Condensate & \makecell{Insulator \\ SPT state \\ Double Condensate \\  Condensate}  & \makecell{$T_x$ \\ $T_x$ \\ $g^\uparrow_x,~g^{\downarrow}_x,~g^\up_x g^\down_x$ \\ $g^\down_x$, $T_x$, $g^\down_x T_x$} &\makecell{SPT-I \\ SPT-II \\ $z$FM-I \\ $z$FM-II}  \\
\hline\hline
\end{tabular}
\label{tab:phases}
\end{table}

In general, parton approaches cannot readily translate chosen states of spinons and chargons to microscopic models. In the present context, we overcome this difficulty by obtaining partons and the gauge field through a sequence of exact lattice duality transformations.

\subsection{Spinons and chargons from dualities}
\label{subsec:threestepduality}

\begin{figure}[h]
    \centering
    \includegraphics[width=0.45\textwidth]{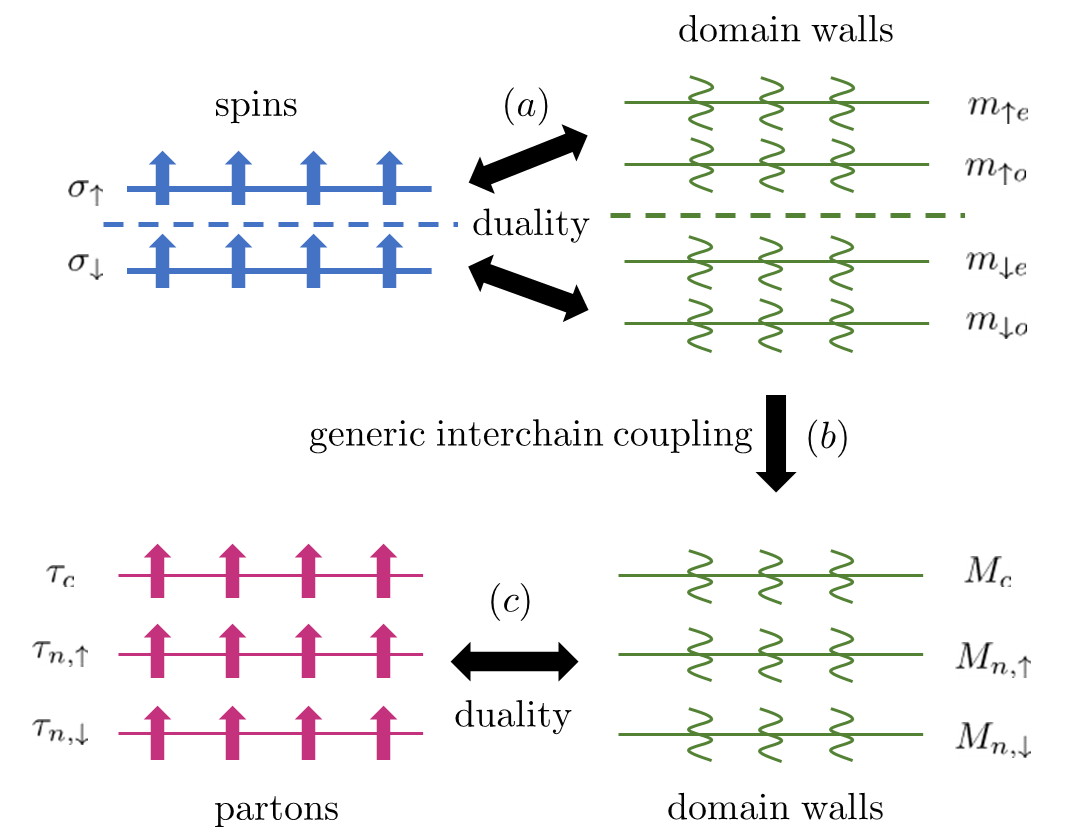}
    \caption{Duality transformations from local spins to partons. $(a)$ In the first step, independent dualities are performed on each leg of the ladder, leading to a chain of two flavors of domain-walls coupled to a $\mathbb{Z}_2$ gauge field on each leg. $(b)$ Generic symmetry-allowed interchain couplings hybridize certain domain-walls, leading to a theory of three domain-walls coupled to two $\mathbb{Z}_2$ gauge fields. $(c)$ Independent dualities on each type of domain-walls result in three flavors of partons coupled to an additional $\mathbb{Z}_2$ gauge field within each flavor. The duality transformations generate overall five flavors of $\mathbb{Z}_2$ gauge fields --- two from the first set of dualities and three from the second. Integrating out four of them leads to a theory of three partons coupled to a single $\mathbb{Z}_2$ gauge field.}
    \label{fig:duality}
\end{figure}

Before deriving the parton gauge theory using dualities on the lattice scale, we discuss the key steps in a more intuitive format. The derivation proceeds in three steps illustrated in Fig.~\ref{fig:duality}. 

\paragraph{Independent dualities for each leg.}

In the first step, the $\uparrow$ and $\downarrow$ legs are independently transformed following Ref.~\onlinecite{JiangMotrunich2019DQCP}. After duality transformations, each leg is described by domain-walls coupled to a $\mathbb{Z}_2$ gauge field. 
Notably, the Spin symmetry in Eq.~(\ref{eqn:Z2sb}) implies that each leg hosts two species of domain walls, $m_{\lambda e}$ and $m_{\lambda o}$, which cannot hybridize and are related by translation symmetry. The actions of $g_z$ and $T_x$ on the domain-walls are listed in Table~\ref{tab:symdomainwall}, while $g^{\lambda}_x$ act trivially on them. Instead, 
the order parameters $\s^z_\l$ are encoded in monopoles $\a^x_\l$ of the $\mathbb{Z}_2$ gauge theory, which are odd under $g^\l_x$. 

Gauge invariant operators consist of even numbers of $m_\l$ for each $\l$ since domain-walls on opposite legs couple to different gauge fields. Specifically, the products $m_{\l \text{e}} m_{\l \text{o}}$ are odd under $g_z$ and even under $T_x$, and thus can be identified as order parameters for the $g_z$-breaking phase. Similarly, $m^2_{\l\text{e}} - m^2_{\l \text{o}}$ are odd under $T_x$ and even under $g_z$, identifying them as order parameters for the VBS phase.  

When none of the domain-walls condenses, the Gauss' constraint induces the condensation of the $\mathbb{Z}^\l_2$ monopole $\langle \a^x_\l \rangle \neq 0$.  
 Consequently, the Charge symmetries $g_x^\lambda$ are broken individually, and the system is in the $z$-FM-II phase of Table~\ref{tab:phases}. When all domain-walls condense $\langle m_{\l \text{e}} \rangle = \pm \langle m_{\l \text{o}} \rangle \neq 0$, the Spin symmetry $g_z$ is broken, and the system is in the $x$-FM phase. When only one of $m_{\l\text{e}},m_{\l\text{o}}$ condenses, each leg forms a VBS; depending on the relative signs of their order parameters, the spin ladder is in the VBS or SPT-I phase, cf.~Section~\ref{subsec:VBS}.

\begin{table}
\centering
\caption{Symmetry transformations of continuum domain-wall variables after a duality transformation of the spin ladder. Monopole operators $\a_\l$ transform nontrivially under Charge symmetries $g^\l_x: \a_\l \to -\a_\l$ and transform trivially under $g_z$ and $T_z$.}
\renewcommand{\arraystretch}{1.2}
\begin{tabular}{ c|cccc| cccc} 
\hline \hline
  & $m_{\uparrow \text{e}}$ &  $m_{\uparrow \text{o}}$ & $m_{\downarrow \text{e}}$ & $m_{\downarrow \text{o}}$  & $\widetilde{m}_{\uparrow 0}$ &  $\widetilde{m}_{\uparrow \pi}$ & $\widetilde{m}_{\downarrow 0}$ & $\widetilde{m}_{\downarrow \pi}$   \\
\hline
$g_z$ & $m_{\uparrow \text{e}}$ & $-m_{\uparrow \text{o}}$ & $-m_{\downarrow \text{e}}$ & $m_{\downarrow \text{o}}$  & $\widetilde{m}_{\uparrow \pi}$ & $\widetilde{m}_{\uparrow 0}$ & $-\widetilde{m}_{\downarrow \pi}$ & $-\widetilde{m}_{\downarrow 0}$ \\
$T_x$ & $m_{\uparrow \text{o}}$ & $m_{\uparrow \text{e}}$ & $m_{\downarrow \text{o}}$ & $m_{\downarrow \text{e}}$ & $\widetilde{m}_{\up 0}$ & $-\widetilde{m}_{\up \pi}$ & $\widetilde{m}_{\down 0}$ & $- \widetilde{m}_{\down \pi}$ \\
\hline\hline
\end{tabular}
\label{tab:symdomainwall}
\end{table}

We note that the Spin symmetry $g_z$ is diagonal in the $m_{\l e},m_{\l o}$ basis, but the lattice translation $T_x$ is not (cf.~Table~\ref{tab:symdomainwall}). Diagonalizing $T_x$ by forming linear combinations 
\begin{equation}
\begin{split}
\widetilde{m}_{\l 0} &= \frac{1}{\sqrt{2}} (m_{\l \text{e}} + m_{\l \text{o}})~, \\
\widetilde{m}_{\l \pi} &= \frac{1}{\sqrt{2}} (m_{\l \text{e}} - m_{\l \text{o}})~,
\end{split} \label{eqn:0piDW}
\end{equation}
results in modes near lattice momenta $0$ and $\pi$. In the alternative representation, the VBS order parameter is $\widetilde{m}_{\l 0} \widetilde{m}_{\l \pi}$, while the order parameter for the $g_z$-breaking phase becomes $\widetilde{m}^2_{\l 0} - \widetilde{m}^2_{\l \pi}$~\cite{JiangMotrunich2019DQCP}.

\paragraph{Interchain interactions.} In the second step, we incorporate inter-chain couplings in the domain-wall representation. Most importantly, the products of all four domain-wall operators, $\m{O}\equiv m_{\up \text{e}} m_{\up \text{o}} m_{\down \text{e}} m_{\down \text{o}}$ or $\widetilde{\m{O}} \equiv \widetilde{m}_{\up 0} \widetilde{m}_{\up \pi} \widetilde{m}_{\down 0} \widetilde{m}_{\down \pi}$, transform trivially under all symmetries. Consequently the terms $\delta H \sim \int_x {\cal O}(x)$ and $\delta \widetilde{H} \sim \int_x \widetilde{\cal O}(x)$ appear in any generic Hamiltonian with interchain couplings, leading to nonzero expectation values for the composite operators. As a result, any \textit{pair} of domain-walls can transform into a complementary pair related by ${\cal O}$ or $\widetilde{\cal O}$. We introduce new domain-wall variables to encode independent degrees of freedom according to
\begin{align}
M_c &= m_{\up\text{e}} m_{\down\text{o}}   &\widetilde{M}_c &= \widetilde{m}_{\uparrow0} \widetilde{m}_{\downarrow\pi} \nonumber\\
&\sim {\cal O} m_{\down\text{e}} m_{\up\text{o}},  &&\sim \widetilde{\cal O}\widetilde{m}_{\down0} \widetilde{m}_{\up\pi},\label{eqn:csdomainwall}\\
M_{n,\lambda} &= m_{\lambda \text{e}},  &\widetilde{M}_{n,\lambda} &= \widetilde{m}_{\lambda 0}, \nonumber
\end{align}
The symmetry transformations of these `charge' domain-walls $M_c$ and `spin' domain-walls $M_{n, \l}$ are listed in Table~\ref{tab:symdomainwall2}. 
Notice that the product of any two new domain-walls $M$ transforms nontrivially under symmetries. Consequently, different $M_a$ cannot hybridize. The `charge' domain-walls couple to the total $\mathbb{Z}_2$ gauge field $\a_\uparrow \a_\downarrow$, while the `spin' domain-walls couple to individual gauge fields $\a_\lambda$. The choice of $M$ or $\widetilde{M}$ amounts to selecting a basis where either $g_z$ or $T_x$ is diagonal. %This completes the second step in Fig.~\ref{fig:duality}. 

\begin{table}[h]
\centering
\caption{Symmetry transformations of continuum domain-wall variables after including interchain couplings. A missing entries indicate that the domain-wall does not have well-defined transformation properties under that symmetry.}
\begingroup
\renewcommand{\arraystretch}{1.5}
\begin{tabular}{ c|ccc|ccc } 
\hline \hline
  & ${M}_c$ &  ${M}_{n, \uparrow}$ & ${M}_{n, \downarrow}$  & $\widetilde{M}_c$ &  $\widetilde{M}_{n,\uparrow}$ & $\widetilde{M}_{ n,\downarrow}$   \\
\hline
$g_z$ & ${M}_c$ & ${M}_{n,\up} $ & $-{M}_{n, \down}$ & $-\widetilde{M}_c$ & %$\widetilde{M}_{n\down} \widetilde{M}_c$ 
&% $-\widetilde{M}_{n \up} \widetilde{M}_c$ 
\\
$T_x$ & ${M}_c$ & %$M_c {M}_{n\down}$ 
&% $M_c {M}_{n\up}$ 
& $-\widetilde{M}_c$ & $\widetilde{M}_{n, \up}$ & $\widetilde{M}_{n, \down}$ \\
\hline \hline
\end{tabular}
\endgroup
\label{tab:symdomainwall2}
\end{table}

\paragraph{Final dualities.}
In the final step, we obtain one chargon $\tau_{c}$ and two spinons $\tau_{n, \lambda}$ by applying an additional duality on each of the three domain-wall types. This resulting parton gauge theory initially contains five $\mathbb{Z}_2$ gauge fields---two from the first and three from the second set of duality transformations. Integrating out four of the five gauge fields yields a theory of three partons coupled to a single $\mathbb{Z}_2$ gauge field $\omega$. The monopole $\omega^x$ of this gauge field depends on the representation of domain-walls. In particular, for partons obtained from $M$, the monopole is the order parameter of the Spin symmetry $g_z$, i.e., $\omega^x \sim\sigma_{x \uparrow} +\sigma_{x \downarrow}$. By contrast, for partons obtained from $\widetilde{M}$, the monopole is odd under translations.% Various combinations of parton states lead to distinct phases shown in Table~\ref{tab:phases}, which we discuss in detail after we perform exact duality transformations on the lattice level.

\begin{table}
\centering
\caption{Mappings of local spins and global symmetries in the parton representation. Chargons and spinons are separately conserved; transmutations between them are forbidden by $g_x^\lambda$ symmetries. However, their quantum numbers are not completely independent. Their total conservation is enforced by a constraint, i.e., a gauged symmetry. $g_z$ symmetry is realized as the conservation of $\mathbb{Z}_2$ gauge flux in the parton representation. (The $\mathbb{Z}_2$ monopole operator violates $g^z$ symmetry.) }
\begin{tabular}{ c c } 
\hline \hline
Spin representation \qquad &\qquad Parton representation   \\
\hline
$\s^z_{\uparrow}$ & $\tau^z_{c} \tau^z_{ n,\uparrow }$ \\
$\s^z_{\downarrow}$ & $\tau^z_{c} \tau^z_{n ,\downarrow }$ \\
$\s^z_{\uparrow} \s^z_{\downarrow}$ & $\tau^z_{n, \uparrow } \tau^z_{n,\downarrow }$ \\
$\s^x_{\uparrow/\downarrow}$ & $\mathbb{Z}_2$ monopole $\omega^x$
\\
$g_x^{\uparrow/\downarrow} = \prod_r \s^x_{\uparrow/\downarrow,r}$ & $g_x^{\uparrow/\downarrow} = \prod_n \tau^x_{n,\uparrow/\downarrow}$
\\
$g_z = \prod_r (\s^z_{\up,r} \s^z_{\down,r})$  &  $\mathbb{Z}_2$ flux $\Phi_\o = \prod \o^z$ \\
\hline \hline
\end{tabular}
\label{tab:partonsym}
\end{table}

We conclude this subsection by discussing the two representations of the domain-walls. The transformation from $(m_{\l \text{e}}, m_{\l \text{o}})$ to $(\widetilde{m}_{\l 0}, \widetilde{m}_{\l \pi})$ cannot be performed on the lattice level. Specifically, linear combinations of real-valued domain-walls in Eq.~(\ref{eqn:0piDW}) are well-defined in the continuum. On the lattice, however, domain-walls are $\mathbb{Z}_2$ variables; linear combinations of $\mathbb{Z}_2$ variables are no longer $\mathbb{Z}_2$ valued. Taking the mod 2 reduction leads to $\mathbb{Z}_2$ variables, but it makes the $\pm$ signs indistinguishable. Therefore, we will opt for the $(m_{\l \text{e}}, m_{\l \text{o}})$ representation when performing lattice-scale manipulations. The dictionary of operators and symmetries in the parton and spin representations are summarized in Table~\ref{tab:partonsym}. We do not include translation symmetry, which is not manifest in the parton representation.
%Notice that the translation symmetry is not manifest with this choice, as is shown in Eq.~(\ref{eqn:csdomainwall}) and Table~\ref{tab:symdomainwall2}. 

\subsection{Exact duality on the lattice}\label{sec.latticeduality}

\begin{table}
\centering
\caption{%\david{Can we add the domain wall expression here} \BH{The nonlocal string in domain wall representation should be represented as a gauge monopole.} 
Mappings between microscopic spins $\s$ and partons $\tau$ coupled with gauge fields $\o$ on the lattice site $r$. In this table, the two species of spinons on the even and odd sublattices $\tau_{n,2r}$ and $\tau_{n,2r+1}$ correspond to $\tau_{n,\l}$ introduced in Section~\ref{subsec:threestepduality}.}
\begin{tabular}{ cc} 
\hline\hline
Local spins & Partons \\
\hline
$\s^z_{\uparrow,2r-1}$ & $\tau^z_{c,2r-1} \tau^z_{n,2r-1}$ \\
$\s^z_{\uparrow,2r}$ & $\tau^z_{c,2r} \o^z_{2\tilde r} \tau^z_{n,2r+1}$ \\
$\s^z_{\downarrow,2r-1}$ &$\tau^z_{c,2r-1} \tau^z_{n,2r}$ \\
$\s^z_{\downarrow,2r}$ & $\tau^z_{c,2r} \tau^z_{n,2r}$ \\
$\s^x_{\uparrow,2r-1}$  & $\o^x_{2\tilde{r}-2} \tau^x_{n,2r-1} $ \\
$\s^x_{\uparrow,2r}$ & $\o^x_{2\tilde r}$ \\
$\s^x_{\downarrow,2r-1}$ & $\tau^x_{c,2r} \tau^x_{n,2r} \o^x_{2\tilde r}$ \\
$\s^x_{\downarrow,2r}$ & $\tau^x_{c,2r} \o^x_{2\tilde r}$ \\
\hline\hline
\end{tabular}
\label{tab:dictionary}
\end{table}

We now perform the dualities outlined above as exact transformations of lattice variables, setting all background gauge fields to unity to lighten the notation. A full list of operators is presented in Table~\ref{tab:dictionary} and derived in Appendix~\ref{app:dictionary}. We illustrate the duality via representative terms that appear in local Hamiltonians with the requisite symmetries. For each term, we provide an expression in microscopic spin variables $\sigma$, in domain-wall variables $\mu$ (the lattice versions of $M_{c,n}$ above), and finally in parton variables $\tau$. Notice that the two species of spinons on the even and odd sublattices $\tau_{n,2r}$ and $\tau_{n,2r+1}$ correspond to $\tau_{n,\uparrow}$ and $\tau_{n,\downarrow}$  introduced in Section~\ref{subsec:threestepduality}.

The unique symmetry-allowed single-site term is the total $\mathbb{Z}_2$ charge density given by
\begin{align}
      Q_r &\equiv \s^x_{\uparrow,r} \s^x_{\downarrow,r} \quad  \quad &&\text{($\mathbb{Z}_2$ Charge density)}\nonumber \\
        &= \mu^z_{c,\tilde r-1}\a^z_{c,r}\mu^z_{c,\tilde r} \quad &&\text{(Domain-wall hopping)}   \nonumber
        \\
              &= \tau^x_{c,r} \quad &&\text{(Chargon density)}   ~,
   \label{eqn:HH1}
\end{align}
where $\tilde{r} = r + 1/2$ denote dual lattice sites. The domain-walls couple to $\mathbb{Z}_2$ gauge fields $\alpha$ living on sites of the original lattice and subject to the Gauss laws

\begin{align}
\a^x_{c,r} \a^x_{c,r+1} =& \begin{cases}\mu^x_{c,\tilde r}\quad &r\text{ even}\\
\mu^x_{c,\tilde r}\mu^x_{n,\tilde r}\quad& r\text{ odd}\end{cases}~, \label{eqn:Gausscs1}
\\
\a^x_{n,r} \a^x_{n,r+1}=& \mu^x_{n,\tilde r}~. \label{eqn:Gausscs2}
\end{align} 
Importantly, the $\mathbb{Z}_2$ charge density is expressible without gauge field in the parton variables (last line) and coincides with the gauge-invariant {\it chargon} density.

Next, the `kinetic' nearest neighbor coupling (analogous to boson hopping) is
\begin{align}
\label{eqn:HH2}
         {\cal O}_{\text{kinetic},r} &\equiv  \s^z_{\uparrow,r} \s^z_{\uparrow,r+1}+ \s^z_{\downarrow,r} \s^z_{\downarrow,r+1} \\
         &= \mu^x_{c,\tilde r} + \mu^x_{c,\tilde r} \mu^x_{n,\tilde r} \nonumber \\
         &=\tau^z_{c,r}\tau^z_{c,r+1}(1+\tau^z_{n,r} \tau^z_{n,r+1})\times \begin{cases}
         1& \text{$r$ odd}~\\
      \omega^z_{\tilde r}\quad& \text{$r$ even}~
     \end{cases}~.\nonumber 
\end{align}
The second line contains a potential energy for the {\it charge} domain-walls and their interaction with the {\it spin} domain-walls. 
The last line describes chargon hopping. It includes a $\mathbb{Z}_2$ gauge field $\omega$ residing on \textit{half} of the dual-lattice sites and satisfying the Gauss law
\begin{equation}
\g^x_{\tilde r} \g^x_{\tilde r+2} =\tau^x_{c,r+1} \tau^x_{c,r+2} \tau^x_{n,r+1} \tau^x_{n,r+2}~,
\label{eqn:gauss}
\end{equation}
with $r$ even. Notice that $\o^x_{2 \tilde{r}}$ is the monopole of the $\mathbb{Z}_2$ gauge field; in an infinite ladder, Eq.~\eqref{eqn:gauss} can be resolved to represent a single $\o^x$ as a string operator
\begin{align}
    \o^x_{2\tilde r} &= \prod_{k=2r+1}^\infty \tau^x_{c,k} \tau^x_{n,k}=
    \sigma^x_{2r,\uparrow}~.
\end{align}

The nearest-neighbor `density-density interaction' within each leg is
\begin{align}
        \m{O}_{\rho \rho,r} =&  \s^x_{\uparrow,r} \s^x_{\uparrow,r+1} + \s^x_{\downarrow,r} \s^x_{\downarrow,r+1} \nonumber \\
        =& \mu^z_{n,\tilde r-1 } \mu^z_{n,\tilde r +1} \left( 1 +  \mu^z_{c,\tilde r-1} \mu^z_{c,\tilde r +1} \right)\nonumber \\&\times \begin{cases}
          \a^z_{c,r} \a^z_{s,r} \a^z_{c,r+1} \a^z_{s,r+1} & \text{ $r$ even} \\
            \a^z_{s,r} \a^z_{s,r+1} & \text{ $r$ odd}
        \end{cases}
        \nonumber \\
        =& \tau^x_{n,r} + \tau^x_{c,r-1} \tau^x_{c,r} \tau^x_{n,r} .
\label{eqn:HH3}
\end{align}
The second line contains a kinetic term for the spin domain-walls, and their interaction with the charge domain-walls. The last line describes a potential energy for the spinons and their interaction with the chargons.

Finally, the on-site symmetries defined in Eqs.~\eqref{eqn:Z2sa},\eqref{eqn:Z2sb} are represented as
\begin{align}
   & g_{x}^{\uparrow}%=\prod_r \sigma^x_{\uparrow,r}
   =\Phi^{(\alpha)}_c \Phi^{(\alpha)}_n =\prod_{r \text{ odd}}\tau^x_{n,r}~,\qquad \quad \nonumber\\
   &g_{x}^{\downarrow}%=\prod_r \sigma^x_{\downarrow,r}
   =  \Phi^{(\alpha)}_n =\prod_{r \text{ even}}\tau^x_{n,r}~, \label{eqn:partonsymmetry}\\
    & g_{z}=
    \prod_{r \text{ even}} \mu^x_{n,\tilde{r}} = \prod_{r \text{ odd}} \mu^x_{n,\tilde{r}} =\Phi_\omega~, \nonumber
\end{align}
where $\Phi^{(\alpha)}_{c(n)} = \prod_r \a^z_{c(n),r}$ and  $\Phi_\omega = \prod_{r}\omega_{2 \tilde r}$ are fluxes of the dynamical gauge fields that couple to domain-walls and partons, respectively. The total charge symmetry can alternatively be expressed as
\begin{equation}
    g_x^{\uparrow}   g_{x}^{\downarrow}=\prod_{r}\tau^x_{c,r}~.
 \label{eqn:partonsymmetry2}
\end{equation}
The lattice parton representation enables the construction of local spin models for all relevant phases. Specifically, known exactly solvable Hamiltonians of insulators, condensates, and SPTs for each parton yields spin models for all the phases in Table~\ref{tab:phases} using the dictionary of Table~\ref{tab:dictionary}. Notice that partons do not transform naturally under individual lattice translation $T_x$ and the reflection $\m{R}$. Instead, they transform naturally under the composite symmetry $\m{R} \times T_x$, up to the dynamical gauge field $\o$. In particular, chargon and spinon density operators transform under $\m{R} \times T_x$ as
\begin{equation}
    \begin{split}
        \tau^x_{c,r} = \s^x_{\up,r} \s^x_{\down,r} &\longrightarrow \s^x_{\down,r+1} \s^x_{\up,r+1} = \tau^x_{c,r+1}, \\
        \tau^x_{n,2r} = \s^x_{\down,2r-1} \s^x_{\down,2r} &\longrightarrow \s^x_{\up,2r} \s^x_{\up,2r+1} = \tau^x_{n,2r+1},
    \end{split}
\end{equation}
Similar transformations hold for chargon and spinon hopping operators $\tau^z_{c(n),r} \tau^z_{c(n),r+1}$.

\section{Phases from parton gauge theories}
\label{sec:partonphase}
In this section, we systematically investigate the phases listed in Table~\ref{tab:phases} via parton gauge theories. To identify the phases, we include the background gauge fields introduced in Section~\ref{sec:couplingbackgroundfield}. We then integrate out chargons and spinons to obtain an effective action of the background $\mathbb{Z}_2$ gauge fields, which diagnoses the phase of the spin ladder. 

\subsection{Induced gauge-field action of chargon and spinon states}
\label{subsec:partongauge}

\subsubsection{Chargon insulators}
\label{sec:CI}

When the chargons are gapped, $\mathbb{Z}_2$ charge fluctuations %\david{fluctuations are always plural} 
are frozen, leading to a {\it charge} insulator. In this case, the induced response of the dynamical gauge field $\o$ is trivial. A specific interaction that leads to a chargon insulator in the atomic limit is
\begin{align}
    H_{\text{CI}} &= -U \sum_r \tau^x_{c,r}~, \label{eqn:ci}
\end{align}
which does not contain couplings to the background fields. Integrating out the chargons amounts to setting $\tau^x_{c,r}=1$, which implies
\begin{align}
    g_x^\uparrow g_x^\downarrow&=1. \label{eqn:ciflux}
\end{align} The Gauss law of Eq.~\eqref{eqn:gauss} with background fields reduces to
\begin{equation}
\tau^x_{n,r+1} \tau^x_{n,r+2}=\g^x_{\tilde r} \g^x_{\tilde r+2} 
\label{eqn:gaussci}
\end{equation}
To fully determine the phase and its broken symmetries, we must further specify which among the four possible states the spinons form; see Section~\ref{subsec:phases}.

\subsubsection{Chargon condensates}
\label{sec:cc}  
Condensation of $\tau^z_c$, which is charged under $\o$ charge, leads to a Higgs phase with a Meissner response, i.e., flux expulsion. An interaction realizing a chargon condensate is 
\begin{equation}
\begin{split}
    H_{\text{CC}} &= -V \sum_{r\text{ even}} \left( \tau^z_{c,r} \O^z_{\text{C}\up, \tilde{r}} \o^z_{\tilde{r}} \tau^z_{c,r+1}  + \tau^z_{c,r-1} \O^z_{\text{C}\down,\tilde{r}-1} \tau^z_{c,r} \right). \label{eqn:cc}
    \end{split}
\end{equation}

Loosely, chargon condensation can be expressed as $\langle \tau^z_{c,r}\rangle  \neq 0$. More precisely, this statement should be understood as
\begin{align}
\lim \limits_{r \rightarrow \infty} \left \langle \tau^z_{c,0}\left(\prod\nolimits_{0<\tilde r'<r} \omega^z_{2 \tilde r}\right)\tau^z_{c,2r}\right\rangle \neq 0~,
\end{align}
since individual $\tau^z_{c,r}$ are not gauge invariant. This correlation function is readily evaluated for the Hamiltonian in Eq.~\eqref{eqn:cc} by noting that all terms commute and have eigenvalues $\pm 1$. 

To obtain the gauge-field response explicitly, we integrate out chargons, which amounts to multiplying all terms in Eq.~(\ref{eqn:cc}). The chargon operators $\tau^z_{c,r}$ cancel and we obtain the Meissner response
\begin{align}
\Phi_\o &\equiv \prod_r \o^z_{2\tilde{r}} = \prod_{r\text{ even}} \O^z_{\text{C}\up,\tilde{r}} \prod_{r\text{ odd}} \O^z_{\text{C}\down,\tilde{r}}~. \label{eqn:omegaflux}
\end{align} One can eliminate the gauge field $\o$ by making a gauge choice such as  $\o^z_{\tilde{r}} = (\O^z_{\text{C}\up,\tilde{r}} \O^z_{\text{C}\down,\tilde{r}-1})$. By contrast, the fluxes of the background gauge fields are unconstrained since $\Omega_\text{S}$ and $\O^z_{\text{C} \up(\down)}$ on odd (even) lattice sites do not enter the Hamiltonian in Eq.~\eqref{eqn:cc}.

\subsubsection{Spinon insulators}\label{pargaraph.spinon}

When the spinons form a conventional (non-topological) insulator, the induced response of the gauge field $\omega$ is trivial. A simple interaction that leads to an atomic spinon insulator is
\begin{align}
 H_\text{SI} = -J_\text{SI}\sum_r \O^z_{\text{S},\tilde{r}-1} \tau^x_{n,r} ~.  \label{eqn:partonsi}
\end{align}
Similar to the chargon insulator, integrating out the spinons amounts to setting $\tau^x_{n,r}= \O^z_{\text{S},\tilde{r}-1}$. By inserting this relation into Eq.~(\ref{eqn:partonsymmetry}), we find that a spinon insulator on the odd (even) lattice sites satisfies
\begin{align}
   g_x^{\up (\downarrow)} = \prod_{r\text{ even (odd)}} \Omega^z_{\text{s},\tilde r} \label{eqn:cicflux}~.
\end{align}
Notice that Eq.~(\ref{eqn:cicflux}) is gauge invariant even though $g^\lambda_x$ depends on the background gauge field. Spinon insulators on both sublattices thus induce a topological response
\begin{align}
g_x^\uparrow g_x^\downarrow & = \Phi_S~. \label{eqn:fluxS}
\end{align}

\subsubsection{Spinon condensates}
When at least one of the two spinons condenses, the gauge field $\o$ is in a Higgs phase. A simple interaction that results in condensation of spinons on the even sublattice is
\begin{align}
    H_{\text{SC}}^\text{even} &= - J_{\text{SC}}^\text{even} \sum_{r\text{ even}}\tau^z_{n,r} \O^z_{\text{C} \up,  \tilde{r}} \O^z_{\text{C}\down,\tilde{r}} \o^z_{\tilde{r}} \tau^z_{n,r+2}\label{eqn:SIC}~.
\end{align}
Integrating out the spinons as we did for chargon condensates leads to the constraint
\begin{equation}
    \begin{split}
        \Phi_\o &= \prod_{r \text{ even}} \O^z_{\text{C}\up,\tilde{r}} \O^z_{\text{C}\down,\tilde{r}}.~
    \end{split} \label{eqn:sicflux}
\end{equation}
For condensation of spinons on the second sublattice, the sum and product over $r$ must be taken over odd sites.
In either case, the flux of $\omega$ is fixed by the background field, i.e., there is a Meissner response.

When both species condense, the even and odd constraints combine into
\begin{equation}
\Phi_{\text{C} \up} \Phi_{\text{C} \down} = \prod_r \O^z_{\text{C} \up, \tilde r} \O^z_{\text{C} \down, \tilde{r}} 
= 1~ \label{eqn:fluxC}~.
\end{equation}
The expulsion of the composite flux $\Phi_{\text{C} \up} \Phi_{\text{C} \down}$ implies that $g_x^{\uparrow}$ and $g_x^{\uparrow}$ are broken, but not necessarily their product; see the discussion at the end of this subsection.

\subsubsection{Spinon SPT states}~
When the spinons form an SPT state protected by the $\mathbb{Z}_2 \times \mathbb{Z}_2$ Charge symmetries, the gauge flux of $\o$ is tied to each of the conserved charges $g^\up_x$ and $g^\down_x$. 
To obtain this topological response explicitly, we place the spinons into the canonical cluster state with Hamiltonian~\cite{RaussendorfBriegel2001cluster} 
     \begin{equation}
     \begin{split}
        H_{\text{SPT}} &= -J_{\text{SPT}} \sum_{r\text{ even}} (\tau^z_{n,r}  \O^z_{\text{C} \up,  \tilde{r}} \O^z_{\text{C} \down,  \tilde{r}} \O^z_{\text{S},\tilde{r}} \o^z_{ \tilde{r}} \tau^x_{n,r+1} \tau^z_{n,r+2} \\
        &\ +  \tau^z_{n,r-1} \tau^x_{n,r} \O^z_{\text{C} \up,  \tilde{r}-1} \O^z_{\text{C} \down,  \tilde{r}-1} \O^z_{\text{S}, \tilde{r}-1} \g^z_{ \tilde{r}} \tau^z_{n,r+1} )~.
     \end{split}
     \label{eqn.hspt.spinon}
    \end{equation}

Integrating out the spinons as before yields two constraints for the dynamical gauge flux, i.e.,
\begin{equation}
\begin{split}
 \Phi_\o &= g_x^\up \left( \prod_{r \text{ even}} \O^z_{\text{C} \up,\tilde{r}} \O^z_{\text{C} \down, \tilde{r}} \O^z_{\text{S}, \tilde{r}} \right)~, \\
 \Phi_\o & = g_x^\down \left( \prod_{r \text{ odd}} \O^z_{\text{C} \up,\tilde{r}} \O^z_{\text{C} \down, \tilde{r}} \O^z_{\text{S}, \tilde{r}} \right)~.
\end{split} \label{eqn:gaugeflux}
\end{equation}
Eliminating the dynamical gauge field $\o$ in Eq.~(\ref{eqn:gaugeflux}) leads to a constraint for the background fluxes
\begin{align}
\Phi_{\text{C} \up} \Phi_{\text{C}\down} \Phi_{\text{S}} &= \prod_r  \O^z_{\text{C}\up,\tilde{r}} \O^z_{\text{C}\down, \tilde{r}} \O^z_{\text{S},\tilde{r}}= g^\uparrow_x g^\downarrow_x~, \label{eqn:fluxCS}
\end{align} 

\subsubsection*{Relations between composite symmetries and fluxes}

The product of different background fluxes can be constrained without fixing individual fluxes [e.g., Eq.~\eqref{eqn:fluxC} and \eqref{eqn:fluxCS}]. Appendix~\ref{append:composite} explains how to determine the broken symmetries in those cases. For the present examples, we redefine Charge and Spin symmetries and the corresponding fluxes according to
\begin{align}
g'^\up_x &= g^\up_x g^\down_x,\quad&& \Phi'_{\text{C}\up} = \Phi_{\text{C}\up}, \nonumber \\
g'^\down_x &= g^\down_x g_z, \quad&& \Phi'_{\text{C}\down} = \Phi_{\text{C}\up} \Phi_{\text{C}\down}, \label{eqn:redefine} \\
g'_z &= g_z, \quad&& \Phi'_{\text{S}} = \Phi_{\text{C}\up} \Phi_{\text{C}\down} \Phi_{\text{S}}. \nonumber 
\end{align}
The constraint in Eq.~\eqref{eqn:fluxC} implies that among the primed symmetries, only $g^{\prime\downarrow}_x$ is broken. Consequently, the original symmetries, $g^{\uparrow}_x,g^{\downarrow}_x$ are both broken while their product as well as $g_z$ are preserved.

\subsection{Phases of the parton gauge theory}
\label{subsec:phases}
Based on the induced actions, we readily identify the phases realized by any combination of chargon and spinons states. The results are summarized in Table~\ref{tab:phases}. 

\subsubsection{$x$-Ferromagnet}
When the chargons and both spinons form insulators, the resulting constraints (\ref{eqn:ciflux}) and (\ref{eqn:fluxS}) imply $\Phi_\text{S}  = 1$, i.e., a Meissner response for the $\mathbb{Z}_2$ Spin background gauge field. Consequently, Spin symmetry $g_z$ is spontaneously broken. The two ground states are the eigenstates of the unconstrained dynamical gauge flux $\Phi_\o = \pm 1$. Their degeneracy implies spontaneously broken flux conservation due to the condensation of a $\mathbb{Z}_2$ monopole~\cite{JiangMotrunich2019DQCP}. The $\mathbb{Z}_2$ monopole is odd under $g_z$, and its condensation breaks this symmetry. 

In the local spin representation, the chargon insulator Hamiltonian of Eq.~\eqref{eqn:ci}, is given by
\begin{align}
H_{\text{CI}} &= - U \sum_r \s^x_{\up,r} \s^x_{\down,r}, \label{eqn:cispin}
\end{align}
identical to the $\mathbb{Z}_2$ charge insulator of Eq.~(\ref{eqn:z2ci}). The spinon insulator Hamiltonian of Eq.~\eqref{eqn:partonsi} is

\begin{equation}
    \begin{split}
         H_{\text{SI}} &= - J_{\text{SI}} \sum_{r \text{ even}} \left( \s^x_{\downarrow,r-1} \O^z_{\text{S},\tilde{r}-1} \s^x_{\downarrow,r} + \s^x_{\uparrow,r} \O^z_{\text{S},\tilde{r}} \s^x_{\uparrow,r+1} \right).
    \end{split} \label{eqn:CISI}
\end{equation}
All terms in $H_{\text{CI}}$ and $H_{\text{SI}}$ commute and the ground state is independent of the coefficients $U,J_{\text{SI}}$. Still, it is convenient to take the limit of infinite $U$, reducing the total Hamiltonian $H$ of spinons and chargons reduces to 
\begin{align}
H |_{U \to \infty} &= -J_{\text{SI}} \sum_r S^x_r \O^z_{\text{S},\tilde{r}} S^x_{r+1}
\end{align}
in the effective spin-1/2 representation of Eq.~\eqref{curly}. Its two degenerate ground states break the Spin symmetry $g_z$. Notice that the ground states for $\Omega^z_S=1$ are translation-invariant although the Hamiltonian $H_\text{SI}$ is not. Consequently, they are also ground states of the translation invariant Hamiltonian $H'= H + T_x H T^{-1}_x$. 

\subsubsection{$y$-Ferromagnet}
When the chargons form insulators, another possibility for spinons is the SPT state. Combining the constraints (\ref{eqn:ciflux}) and (\ref{eqn:fluxCS}), we obtain a Meissner response $\Phi'_{\text{S}} = 1$, which implies broken $g'_z$. As a result, all Charge and Spin symmetries $g^\up_x$, $g^\down_x$, and $g_z$ are broken individually, and any product of two of them is conserved; see Eq.~\eqref{eqn:redefine}.  In the local spin representation, the spinon SPT Hamiltonian becomes
\begin{equation}
    \begin{split}
          H_{\text{SPT}} = &-J_{\text{SPT}}  \sum_{r\text{ odd}} \s^z_{\uparrow,r}  \s^z_{\uparrow,r+1} \s^y_{\downarrow,r}  \O'^z_{\text{S},\tilde{r}} \s^y_{\downarrow,r+1}   \\
    & -J_{\text{SPT}}  \sum_{r\text{ even}}\s^z_{\downarrow,r}  \s^z_{\downarrow,r+1} \s^y_{\uparrow,r} 
     \O'^z_{\text{S},\tilde{r}}\s^y_{\uparrow,r+1},
    \end{split} \label{eqn:CISSPT}
\end{equation}
where $\O'^z_{\text{S},\tilde{r}}$ is the $\mathbb{Z}_2$ gauge field for $g'_z$. 

As before, spinon and chargon terms commute, and it is expedient to take the $U\rightarrow \infty$ limit to obtain
\begin{equation}
  H |_{U \to \infty} = - J_{\text{SPT}} \sum_r S^y_r \O'^z_{\text{S},\tilde{r}} S^y_{r+1}~.
\label{eqn.57}
\end{equation}
Its ground state has $\langle S^y_r \rangle \neq 0$, which spontaneously breaks $g^\uparrow_x,g^\downarrow_x$ and $g_z$ individually but preserves their products according to Eq.~\eqref{eqn:Z2seff}.

As for the $x$FM phase discussed above, the ground states are translation symmetric (for uniform $\Omega^{\prime z}_\text{S}$) even though the Hamiltonian is not. As a result, the translation-invariant Hamiltonian $H' = H + T_x H T_x^{-1}$ has the same ground states as $H$.

\subsubsection{$z$-Ferromagnet}\label{phases.zfm}
When the chargons are insulating and both spinons condense, the induced spinon response (\ref{eqn:fluxC}) implies spontaneously broken  $g^\up_x, g^\down_x$ symmetries. The chargon response does not impose any additional constraints. Consequently, all other symmetries are preserved, in particular the product $g^\up_xg^\down_x$.

In the local spin representation, the Hamiltonian for spinon double condensates becomes
\begin{equation}
        H_{\text{SC}}  = - J_{\text{SC}} \sum_r \s^z_{\uparrow,r}  
        \O^z_{\text{C}\up,\tilde{r}} \s^z_{\uparrow,r+1} 
    \s^z_{\downarrow,r}
 \O^z_{\text{C}\down,\tilde{r}}   \s^z_{\downarrow,r+1}.
   \label{eqn:CISC}
\end{equation}
In light of the commutation between all terms in $H_{\text{SC}}$ and $H_\text{CI}$ we again take the infinite $U$ limit to obtain
\begin{equation}
H |_{U \to \infty} = - J_{\text{SC}} \sum_r S^z_r \O^z_{\text{C}\up,\tilde{r}} \O^z_{\text{C}\down,\tilde{r}}S^z_{r+1}.
\end{equation}
The reduced Hamiltonian describes a ferromagnet along the $z$-direction. Here, both the Hamiltonian and the ground states are translation-invariant for uniform $\Omega_C^\lambda$.

\subsubsection{Valence bond solid}
A fourth phase arises when the chargon and the $\tau_{n,\text{odd}}$ spinons form insulators while $\tau_{n,\text{even}}$ spinons condense (the opposite choice for the spinons is related by translation symmetry). The induced responses (\ref{eqn:ciflux}), \eqref{eqn:cicflux} and (\ref{eqn:sicflux}) imply a Meissner response for the dynamical gauge field but do not constrain any background fluxes. Consequently, the ground state is unique and preserves all symmetries.

The Hamiltonian for the spinon single condensate is
\begin{align}
        H_{\text{VBS}} &= H^{\text{even}}_{\text{SC}} + H^{\text{odd}}_{\text{SI}}\label{eqn:spinonvbs} \\
        &= -J^{\text{even}}_{\text{SC}} \sum_{r \text{ even}} \tau^z_{n,r} \o^z_{\tilde{r}} \tau^z_{n,r+2} - J^{\text{odd}}_{\text{SI}} \sum_{r \text{ odd}} \tau^x_{n,r} \nonumber\\
        &= -J^{\text{even}}_{\text{SC}} \sum_{r \text{ even}} \s^z_{\up,r} \s^z_{\up,r+1} \s^z_{\down,r} \s^z_{\down,r+1}\nonumber \\
        & \quad - J^{\text{odd}}_{\text{SI}} \sum_{r \text{ odd}} \s^x_{\up,r-1} \s^x_{\up,r}.\nonumber
\end{align}
All terms in the total Hamiltonian $H = H_{\text{CI}} + H_{\text{VBS}}$ commute. Taking the $U \to \infty$ limit again, we obtain
\begin{equation}
\begin{split}
H |_{U \to \infty} =& -J_\text{VBS}^x \sum_{r 
 \text{ even}} S^x_r \O^z_{\text{S},\tilde{r}} S^x_{r+1} \\&-J_\text{VBS}^z \sum_{r 
 \text{ even}} S^z_r \O^{\prime z}_{\text{C} \down,\tilde{r}} S^z_{r+1}
 \end{split}
 \label{eqn.vbs.S}
\end{equation}
Here both the Hamiltonian and its ground state break translation invariance. The translation invariant Hamiltonian $H' = H + T_x H T_x^{-1}$ realizes a different phase, i.e., it forms the $x$FM or $z$FM, depending on which coupling is larger. We will return to this point in the next section, where we discuss translation invariant models.

\subsubsection{Topological states (SPT-I and SPT-II)}
\label{subsec:topological}
When the chargons condense, the dynamical gauge field is in a Higgs phase, and the spinons are promoted to physical variables, see Table~\ref{tab:dictionary}. Both the trivial insulator and the cluster state of spinons carry a topological response, Eqs.~\eqref{eqn:fluxS} and \eqref{eqn:fluxCS}, and preserve all onsite symmetries. Consequently, the spin ladder forms distinct topological states for the two cases.

The microscopic Hamiltonian for the SPT-I phase is the sum of Eq.~\eqref{eqn:CISI} and 
\begin{equation}
H_{\text{CC}} = -V \sum_{r\text{ even}} \left( \s^z_{\up,r} \O^z_{\text{C}\up,\tilde{r}} \s^z_{\up,r+1}+\s^z_{\down,r-1} \O^z_{\text{C}\down, \tilde{r}-1} \s^z_{\down,r} \right).\label{eqn:ccspin}
\end{equation}
We already discussed the resulting ground state in Fig.~\ref{fig:VBSSPT}(d) and computed its partition function in Table~\ref{tab:pf}. Translation invariance is broken by the Hamiltonian and its ground states. The translation invariant Hamiltonian $H' = H + T_x H T^{-1}_x$ leads to either the $x$FM or $z$FM-I phase, depending on which of $J_{\text{SI}}$ and $V$ is larger. We will discuss this point in more detail in the next section.

The microscopic Hamiltonian for the SPT-II phase is the sum of Eqs.~\eqref{eqn:CISSPT} and \eqref{eqn:ccspin}. To analyze it using local degrees of freedom, we represent the ground state manifold of the dimerized Hamiltonian in Eq.~\eqref{eqn:ccspin} by an effective local spin-1/2 for each dimer. For uniform $\Omega^z_{C,\lambda,r}$ this representation is given by
\begin{align}
&\sigma^z_{\uparrow,2r} \text{ and }\sigma^z_{\uparrow,2r+1} \rightarrow \mathsf{S}^z_{2\tilde r}\quad&&\sigma^{y}_{\uparrow,2r} \sigma^y_{\uparrow,2r+1} \rightarrow \mathsf{S}^x_{2\tilde r}~,\nonumber\\
&\sigma^z_{\downarrow,2r}\text{ and }\sigma^z_{\downarrow,2r-1}  \rightarrow \mathsf{S}^z_{2\tilde r-1}\quad&&\sigma^y_{\downarrow,2r}\sigma^y_{\downarrow,2r-1}  \rightarrow \mathsf{S}^x_{2\tilde r-1}~. \label{eqn:Trep}
\end{align}
Projecting Eq.~(\ref{eqn:CISSPT}) into this manifold yields the canonical cluster state Hamiltonian
\begin{equation}
          H_{\text{SPT}} \rightarrow -J_{\text{SPT}}  \sum_{r} \mathsf{S}^z_{\tilde r-1} \mathsf{S}^x_{\tilde r } \O'^z_{\text{S},\tilde{r}} \mathsf{S}^z_{\tilde r+1} 
\end{equation}

Both the Hamiltonian and its ground state break translation symmetry. Restoring translation invariance by taking $H' = H + T_x H T_x^{-1}$ leads to either the $y$FM or $z$FM-I phase, depending on which coupling is larger. We will return to this point in the next section. 
Notice that this phase is distinct from all the four phases shown in Fig.~\ref{fig:VBSSPT}, which follows from its background gauge response and alternatively from the partition functions listed in Table~\ref{tab:pf2} below.

\begin{table}[h!]
\centering
\caption{Partition functions $Z[g,h^*]$ for the four distinct topological phases in Fig.~\ref{fig:VBSSPT} and the SPT-II phase with a twisted boundary condition characterized by $h^*=(1,-1,1)$. Notice that Table~\ref{tab:pf} lists the partition functions for a different boundary condition that discriminates the four VBS states there. }
\begin{tabular}{ c|ccc } 
\hline \hline
$g$ & $(a),(d)$ & $(b),(c)$ & SPT-II \\
\hline
$g^\up_x$ &+1  & +1 & $-$1\\
$g^\down_x$ &+1 & +1 & +1\\
$g_z$ & +1 & $-$1 & +1\\
\hline \hline
\end{tabular}
\label{tab:pf2}
\end{table}

\subsubsection{Additional $z$-Ferromagnets (zFM-I and zFM-II)} 
\label{subsec:morezFM}

The final two possibilities are given by a chargon condensate while one or both spinons also condense. As in Section~\ref{subsec:topological}, the condensation of chargons leads to a Higgs phase of the gauge field $\o$. 

When both species of spinons condense, the constraints~(\ref{eqn:omegaflux}), (\ref{eqn:sicflux}), and (\ref{eqn:fluxC}) lead to individual Meissner responses $\Phi_{\text{C}\up}= \Phi_{\text{C} \down} = 1$, indicating that both Charge symmetries $g^\up_x$ and $g^\down_x$ are broken separately, which leads to the $z$FM-I phase. In contrast to the $z$FM phase of Section~\ref{phases.zfm}, the total charge symmetry $g^\up_x g^\down_x$ is also broken here. In the local spin representation, the Hamiltonian $H = H_{\text{CC}} + H_{\text{SC}}$ describes two decoupled ferromagnetic Ising chains, which break $g^\up_x$ and $g^\down_x$ respectively.

Notice that the ground states are translation invariant and can thus be realized by $H' = H + T_x H T^{-1}_x$. The LSM constraints permit such a translation symmetric ground state when both $g^\lambda_x$ are broken, as discussed at the end of Section~\ref{sec:symmetries}. 

When only one species of spinons condenses, the constraints (\ref{eqn:omegaflux}) and (\ref{eqn:sicflux}) lead to the Meissner response $\Phi_{\text{C}\down} = 1$, which implies that $g^\down_x$ is broken. In the local spin representation, the ground state of $H = H_{\text{CC}} + H_{\text{VBS}}$ [from Eq.~(\ref{eqn:spinonvbs}) and (\ref{eqn:ccspin})] is a VBS state in the upper leg and a $z$-ferromagnet in the lower leg. As a result, $g^\down_x$, $T_x$ and their product $g^\down_x T_x$ are all broken, i.e., the spin ladder is in the $z$FM-II phase. Symmetrizing with respect to translations does not affect the lower leg, but places the upper leg into an $x$ and $z$ ferromagnet depending on the relative magnitudes of $J^{\text{odd}}_{\text{SI}}$ and $V + J^{\text{even}}_{\text{SC}}$.

\section{Translation invariant Hamiltonians}
\label{sec:partontranslation}

As explained in Section~\ref{sec.latticeduality}, the exact dualities we derive do not keep translation symmetry manifest. Consequently, most of the microscopic Hamiltonians constructed in Section~\ref{sec:partonphase} also do not respect this symmetry. However, for phases that break either $g_z$ or both $g_x^\uparrow,g_z^\downarrow$, a translation symmetric phase is not precluded by LSM constraints. Indeed, for those phases, a translation invariant parent Hamiltonian can be readily obtained by symmetrizing according to $H'= H + T_x H T^{-1}_x$. 

In contrast, for the VBS, SPT-I, SPT-II, and $z$FM-II phases, LSM constraints imply that translation symmetry cannot be restored without rendering the system critical or breaking (additional) onsite symmetries. Still, the same ground states can arise from a translation-invariant Hamiltonian by spontaneously breaking translation symmetry. Since parton variables obscure the role of translations, we use microscopic spin variables to demonstrate this effect and suppress background fields.

\subsubsection{Valence bond solid}
\label{subsec:vbs}
Symmetrizing the Hamiltonian in Eq.~\eqref{eqn.vbs.S} results in
\begin{align}
    H'_\text{VBS} &= -J^x_{\text{VBS}} \sum_r S^x_r S^x_{r+1} - J^z_{\text{VBS}} \sum_r S^z_r S^z_{r+1}., \label{eqn:effXYchain}
\end{align}
which realizes the $x$FM for $J^x_{\text{VBS}} > J^z_{\text{VBS}}$ and the $z$FM for $J^x_{\text{VBS}} < J^z_{\text{VBS}}$. At the critical point $J^x_{\text{VBS}} = J^z_{\text{VBS}}$, the Hamiltonian describes a gapless XY model. Second neighbor couplings 
\begin{align}
    H_{\text{NNN}} &= \frac{1}{2} \sum_r \left( K^x_{\text{VBS}} S^x_r S^x_{r+2} + K^z_{\text{VBS}} S^z_r S^z_{r+2} \right)
\end{align}
frustrate both orders and instead drive the system into a VBS phase. In particular, the two degenerate VBS states related by translation symmetry are exact ground states of $ H'_\text{VBS} + H_{\text{NNN}}$ 
with $J^{x(z)}_{\text{VBS}}=K^{x(z)}_{\text{VBS}}$~\cite{MajumdarGhosh1969nnn}.

\subsubsection{Topological states (SPT-I and SPT-II)}
\label{subsec:SPT}
~Symmetrizing the parent Hamiltonian of the SPT-I state, $H = H_{\text{CC}} + H_{\text{SI}}$, yields
\begin{align}
    H' &= \sum_{r,\l = \uparrow,\downarrow} (-V \s^z_{\l,r} \s^z_{\l,r+1} - J_{\text{SI}} \s^x_{\l,r} \s^x_{\l,r+1}), %\label{eqn:LL1spin}
\end{align}
which is identical to two copies of Eq.~(\ref{eqn:effXYchain}). Taking $V>J_{\text{SI}}$ leads to the $z$FM-I phase, while taking $V<J_{\text{SI}}$ leads to the $x$FM phase.  At the transition, $V=J_{\text{SI}}$, the Hamiltonian $H'$ describes separate XY models for each leg. As discussed above for the VBS state, second-neighbor couplings frustrate both orders and promote a VBS on each leg, leading to the four phases shown in Fig.~\ref{fig:VBSSPT}. Adding weak interchain couplings
\begin{align}
H_\text{int}&= J_\text{int} \sum_r \s^z_{\up,r} \s^z_{\up,r+1} \s^z_{\down,r} \s^z_{\down,r+1}    
\end{align}
reduces the degeneracy from four to two and stabilizes the SPT-I phase with spontaneously broken translation symmetry.  

Symmetrizing the Hamiltonian of the SPT-II state $H = H_{\text{CC}} + H_{\text{SPT}}$  yields
\begin{equation}
\begin{split}
   H' &= -J_{\text{C}} \sum_{r, \l = \up, \down}  \s^z_{\l,r} \s^z_{\l,r+1} \\
   & \quad - J_{\text{S}} \sum_{r,\l = \up,\down} \s^z_{\l,r} \s^z_{\l,r+1} \s^y_{-\l,r} \s^y_{-\l,r+1}.
\end{split} \label{eqn:Txccsspt}
\end{equation}
Here, $J_C > J_S$ realizes the $z$FM-I, and $J_S>J_C$ the $y$FM. The transition between these phases is revealed by a duality transformation (see Appendix~\ref{appendix:cssspt}) mapping Eq.~\eqref{eqn:Txccsspt} onto
\begin{equation}
\begin{split}
    \tilde H &= \sum_{a = A,B} \left( -J^a_{\text{C}} \sum_r \m{Z}^a_r \m{Z}^a_{r+1} - J^a_{\text{S}} \sum_r \m{X}^a_r \m{X}^a_{r+1} \right).
\end{split} \label{eqn:dualbasis}
\end{equation}
Translation symmetry interchanges $A,B$ and enforces $J_C^A = J_C^B$ and  $J_S^A = J_S^B$. There is then a single phase transition between the two ordered states when all four couplings are equal.

To identify the SPT-II, we observe that the unprimed Hamiltonian $H$ maps onto $ \tilde H$ with $J_C^A=J_S^B =0$, which implies $\langle {\cal Z}_A\rangle \neq 0$ and $\langle {\cal X}_B\rangle \neq 0$. To access the same phase without breaking the $A,B$ symmetry explicitly, one can perturb the critical point of $\tilde H$ by
\begin{align}
    \tilde H_\text{int} &=\delta \sum_r ( \m{Z}^A_r \m{Z}^A_{r+1} - \m{X}^A_r \m{X}^A_{r+1})(\m{Z}^B_r \m{Z}^B_{r+1} -\m{X}^B_r \m{X}^B_{r+1}) \label{eqn:intVBS},
\end{align}
which represents a marginally relevant perturbation. Translating the model $\tilde H|_{J_C=J_S} + \tilde H_\text{Int}$ back to microscopic spins yields a Hamiltonian [Eq.~(\ref{eqn:localHint})] whose ground state spontaneously breaks translation symmetry and realizes the SPT-II phase. 

\subsubsection{zFM-II}
Symmetrizing the Hamiltonian of the $z$FM-II state, $H = H_{\text{CC}}+ H^{\text{even}}_{\text{SC}}+ H^{\text{odd}}_{\text{SI}}$, leads to
\begin{equation}
\begin{split}
    H' &= -V \sum_{r,\l=\up,\down} \s^z_{\l,r} \s^z_{\l,r+1} - J^{\text{odd}}_{\text{SI}} \sum_r \s^x_{\up,r} \s^x_{\up,r+1} \\
    & \quad - J^{\text{even}}_{\text{SC}} \sum_r \s^z_{\up,r} \s^z_{\up,r+1} \s^z_{\down,r} \s^z_{\down,r+1}.
    \end{split}
\end{equation}
The $\sigma^z$ operators on the lower legs are conserved quantities and satisfy $\s^z_{\down,r} \s^z_{\down,r+1}=1$ in the ground state.  Consequently, the Hamiltonian $H'$ reduces to an XY chain for the upper leg, whose critical point occurs for $J^\text{odd} = V + J_\text{SC}^\text{even}$.  Adding next-neighbor couplings can drive the upper leg into a VBS, spontaneously breaking translation symmetry. For one sign of this single-leg VBS order parameter, the ladder realizes the same state described in \ref{subsec:morezFM}.

\section{Phase transitions}
\label{sec:transition}
Having investigated all relevant phases, we turn to their phase transitions. We begin with the microscopic models derived in Section~\ref{sec:partonphase} and address translation-invariant models in Section~\ref{subsec:tt}.

\subsection{Chargon transitions}
\label{subsec:citocc}

The dynamical $\mathbb{Z}_2$ gauge field in one dimension can be reduced to a boundary condition. Consequently, we expect critical properties to be accurately captured by a mean-field analysis that ignores the dynamical gauge field and treats chargons as microscopic variables. The insulator and condensate of chargons are distinct by a single $\mathbb{Z}_2$ order parameter, which implies an Ising transition with central charge $c=\frac{1}{2}$.

We confirm this finding for a chargon transition in spinon double condensates using the local spin variables of Eq.~\eqref{eqn:dictSS}. The effective Hamiltonian
\begin{align}
   H = -\sum_r(J_\text{CI} \tilde{S}^x_r +J_\text{CC} \tilde{S}^z_r \tilde{S}^z_{r+1})
\end{align}
describes a transverse field Ising model, paradigmatic for $c=\frac{1}{2}$ criticality.
\subsection{Spinon transitions}
\label{subsec:spinontransition}
As for the chargons above, we treat the spinons as microscopic variables. The trivial insulator differs from the single condensate by one $\mathbb{Z}_2$ order parameter, and their transition is of the Ising type with central charge $c=\frac{1}{2}$. Likewise, the transition between single and double condensates has $c=\frac{1}{2}$. A direct transition between trivial insulator and double condensate is consequently of the XY type with $c=1$. The cluster state is also separated from the trivial insulator and double condensate by $c=1$ transitions, and from the single condensate by an Ising transition~\cite{Verresen2017spt1d}. 

We confirm these results by an analysis in the local effective spin representation of Eq.~(\ref{curly}) for chargon insulators, where we find
\begin{align}
H_\text{eff} &=  -\sum_r (J_{\text{SI}} S^x_r S^x_{r+1} + J_{\text{SC}} S^z_r S^z_{r+1} + J_{\text{SPT}} S^y_r S^y_{r+1})\nonumber\\
&\quad -J_{\text{single}}\sum_{r\text{ even}} ( S^x_r S^x_{r+1}+S^z_r S^z_{r+1})~.
\end{align} In particular, transitions between phases described by the first line ($J_{\text{single}}=0$) occur when the largest two coupling constants coincide. There, Hamiltonian exhibits a $U(1)$ symmetry and the transition is of the XY type. Likewise, one readily confirms that the remaining transitions are of the Ising type.

\subsection{Translation enhanced phase transitions}
\label{subsec:tt}

\label{subsec:Txenhanced}
In Secs.~\ref{sec:partonphase} and~\ref{sec:partontranslation}, we discussed how all eight phases can arise in translation-invariant spin ladders. We do not expect the Hamiltonian's translation invariance to affect transitions between phases agnostic to this symmetry, i.e., those permitted to be translation-invariant by LSM. 
Likewise, transitions between phases in which translation symmetry breaking is enforced by the dogma of LSM are not expected to depend on the Hamiltonian' translation invariance. The most interesting cases occur at transitions between agnostic and dogmatic phases.

\begin{figure}[h]
    \centering
    \includegraphics[width=0.5\textwidth]{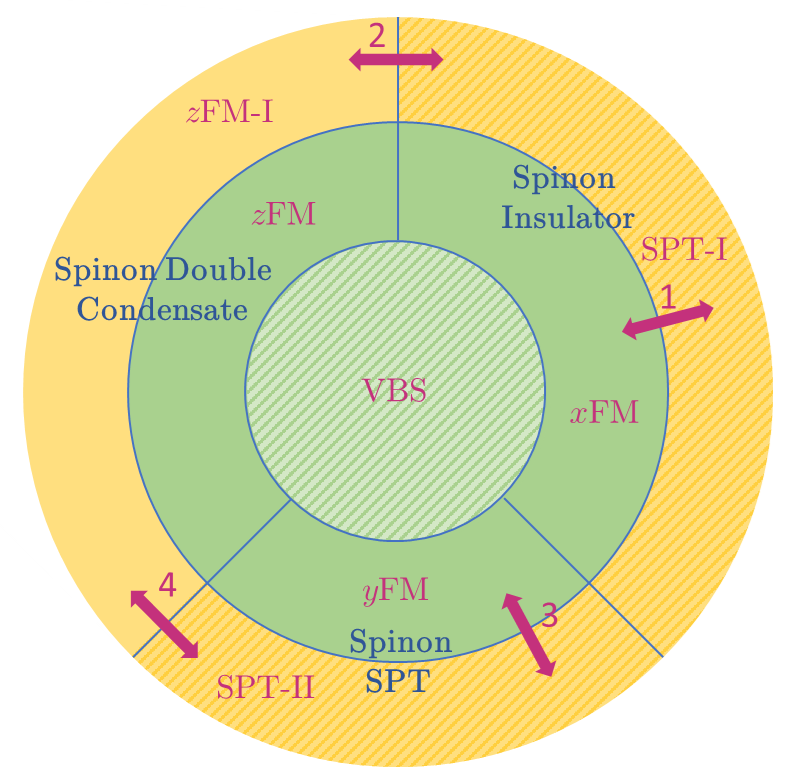}
    \caption{Translation-symmetry enhanced phase transitions. The green and yellow annuli represent chargon insulators and chargon condensates, respectively. All phases with shaded lines break lattice translation symmetry spontaneously. We do not include transitions that involve the $z$FM-II phase, which breaks the reflection symmetry $\m{R}$.}
    \label{fig:translation}
\end{figure}

We restrict ourselves to phases compatible with the composite symmetry $\m{R} \times T_x$, which was discussed at the beginning of Section~\ref{sec.latticeduality}. As shown in the phase diagram of Fig.~\ref{fig:translation}, four transitions are potentially affected by translation symmetry. In this section, we discuss phase transitions 1 and 2 in Fig.~\ref{fig:translation} and leave the discussions for transitions 3 and 4 to Appendix~\ref{appendix:transition}, as their analyses mirror the first two.

\subsubsection{Transition between $x$FM and SPT-I}Transition 1 in Fig.~\ref{fig:translation} connects the $x$FM phase (chargon insulator, spinon insulator) to the SPT-I phase (chargon condensate, spinon insulator). Both phases can be accessed by perturbing the gapless parent Hamiltonian
\begin{align}
    H_\text{p} &=- \sum_{r,\l = \uparrow,\downarrow} (\s^z_{\l,r} \s^z_{\l,r+1} + \s^x_{\l,r} \s^x_{\l,r+1}-\Delta\s^y_{\l,r} \s^y_{\l,r+1}).~ \label{eqn:LL1spin}
\end{align}
Specifically, $x$FM arises upon perturbing $H_\text{p}$ with $\sigma^x_{\uparrow}\sigma^x_{\downarrow}$, and SPT-I  as prescribed in Section~\ref{sec:partontranslation}. To determine the critical properties of the phase transition, we employ Abelian bosonization~\cite{cftbook,Giamarchibook,stone1994bosonization,Shankar1995bosonization,Gogolin:2004rp}. The lattice model $H_\text{p}$ with $\Delta \in (-1,1]$
corresponds to the Hamiltonian density 
\begin{equation}
{\cal H}_{\text{LL}} = \frac{v_F}{2\pi}\sum_{\l=\uparrow,\downarrow}[K_\l^{-1}(\partial_x \varphi_\l)^2+K_\l(\partial_x \theta_\l)^2], \label{eqn:LL1}
\end{equation}
with $K_\lambda \geq \frac{1}{2}$. 
The long-wavelength variables $\varphi,\theta$ satisfy the canonical commutation relations 
\begin{align}
    \left[ \p_x \t_\l(x), \varphi_{\l'}(x') \right] &= i \pi \d_{\l \l'} \d(x-x'), \quad \l,\l' = \up,\down,\label{eqn:canonical}
\end{align}
and encode local spin operators via 
\begin{equation}
\begin{split}
    \s^x_{\l,r} + i \s^z_{\l,r} &\sim e^{-i \varphi_\l} \left[ C_1 + C_2 (-1)^r \cos{2\t_\l} \right], \\
    \s^y_{\l,r} &\sim \frac{2\p_x \t_\l}{\pi} + C_3 (-1)^r \sin{2\t_\l}, ~,
\end{split}
\end{equation}
where $C_{1,2,3}$ are normalization factors. The transformations of the bosonized variables under symmetries are listed in Table~\ref{tab:bosonized}.

A sine-Gordon term leading to the $x$FM phase is 
\begin{align}
\d \m{H}_{x\text{FM}} &= - 2 J_x \cos{\varphi_\uparrow} \cos{\varphi_\downarrow},
\end{align}
whose ground states exhibit $\langle \cos{\varphi_\uparrow}\rangle =\langle \cos{\varphi_\downarrow}\rangle \neq 0$, spontaneously breaking the Spin symmetry $g_z$. The SPT-I phase arises from the perturbation
\begin{equation}
  \d \m{H}_{\text{SPT-I}} =+2 J_{\text{SPT-I}} \cos{(2\t_\uparrow)} \cos{(2\t_\downarrow)}, \label{eqn:sg}
\end{equation}
realizing a phase with $\langle\cos{(2\t_\uparrow)}\rangle =-\langle \cos{(2\t_\downarrow)}\rangle \neq 0$, which spontaneously breaks translation invariance.  

\begin{table}[h]
\centering
\caption{Symmetry transformations of bosonized variables in the low-energy effective theory.}
\begingroup
\renewcommand{\arraystretch}{1.2}
\begin{tabular}{ c|cccc } 
\hline \hline
  & $\varphi_\up$ &  $\t_\up$ & $\varphi_\down$  & $\t_\down$    \\
\hline
$g^\up_x$ & $-\varphi_\up$ & $-\t_\up$ & $\varphi_\down$ & $\t_\down$ \\
$g^\down_x$ & $\varphi_\up$ & $\t_\up$ & $-\varphi_\down$ & $-\t_\down$ \\
$g_z$ & $-\varphi_\up+\pi$ & $-\t_\up$ & $-\varphi_\down+\pi$ & $-\t_\down$ \\
$T_x$ & $\varphi_\up$ & $\t_\up + \frac{\pi}{2}$ & $\varphi_\down$ & $\t_\down+ \frac{\pi}{2}$ \\
\hline \hline
\end{tabular}
\endgroup
\label{tab:bosonized}
\end{table}

To understand the transition between $x$FM and SPT-I phases, we perform a canonical transformation to new variables  $\varphi_\pm = \frac{1}{2}(\varphi_\up \pm \varphi_\down)$ and $\t_\pm = \t_\up \pm \t_\down$, which preserve the commutation relations of Eq.~(\ref{eqn:canonical}). The transition is then described by 
\begin{equation}
    \begin{split}
        \m{H}^{(1)}_{\text{transition}} &= \m{H}_{\text{LL}} + \d H_{x\text{FM}} + \d H_{\text{SPT-I}}\\
        &=\m{H}_{\text{LL}} +\sum_{\kappa=\pm}\left( J_{\text{SPT-I}} \cos{2\t_\kappa} - J_x\cos{2\varphi_\kappa} \right).
    \end{split}
    \label{eqn.xfmspt1}
\end{equation}
The second line describes independent Ising transitions for $\kappa=\pm$~\cite{LecheminantGogolinNersesyan2002selfdual}. In particular, 
the Ising disorder parameters are $\mu_\pm \sim \cos{\t_\pm}$. More transparently, we can fermionize $\varphi_\pm, \t_\pm$
with Majorana representations as

\begin{equation}
    \begin{split}
        \b{\xi}_\pm &= \begin{bmatrix}
            \cos{(\varphi_\pm + \t_\pm)} \\
            \cos{(\varphi_\pm - \t_\pm)}
            \end{bmatrix},
            \ \b{\eta}_\pm = \begin{bmatrix}
                \sin{(\varphi_\pm + \t_\pm)} \\
            \sin{(\varphi_\pm - \t_\pm)}
            \end{bmatrix}.
        \end{split}
\end{equation}
Notice that each doublet contains one left-moving and one right-moving Majorana fermion.  The Luttinger liquid Hamiltonian of Eq.~(\ref{eqn:LL1}) with $K_\up = K_\down = 2$ maps onto the non-interacting Majorana fermion Hamiltonian
\begin{equation}
    {\cal H}_{\text{kin}} = i v_F \sum_{\kappa = \pm} \left( \b{\xi}_\kappa \tau^z \p_x \b{\xi}_\kappa + \b{\eta}_\kappa \tau^z \p_x \b{\eta}_\kappa \right),
\end{equation}
where $\tau^z$ is the Pauli-$z$ matrix. The sine-Gordon terms in Eq.~(\ref{eqn.xfmspt1}) correspond to mass terms for the Majorana fermions, i.e.,

\begin{equation}
    \begin{split}
        \m{H}_{\text{mass}} &= (J_{\text{SPT-I}} - J_x) \sum_{\kappa = \pm} i \b{\xi}_\kappa \tau^x \b{\xi}_\kappa \\
        &\quad + (J_{\text{SPT-I}} + J_x) \sum_{\kappa = \pm} i \b{\eta}_\kappa \tau^x \b{\eta}_\kappa.
    \end{split} \label{eqn:Hint1Majorana}
\end{equation}
 The critical points of ${\cal H}_{\text{kin}}+ \m{H}_{\text{mass}}$ occur at $J_{\text{SPT-I}} = \pm J_x$ and describe two massless Majorana fermions with total central charge $c=\frac{1}{2}+\frac{1}{2}=1$.

Allowing $K_\up = K_\downarrow$ to deviate from $2$ introduces interactions among Majorana fermions with equal $\kappa$, which are irrelevant at the transition. Higher-order nonlinear terms such as $\cos 4 \theta_\uparrow + \cos 4 \theta_\downarrow \sim \cos 2 \theta_+ \cos 2 \theta_-$ correspond to marginal four-fermion interactions between Majorana fermions with different $\kappa$, which modify the critical behavior from Ising $\times$ Ising to a generic point on the Ashkin-Teller (AT) line without changing the central charge~\cite{AshkinTeller1943,Ginsparg1988cft}. Finally, a direct coupling between the two Ising disorder parameters $~\mu_+ \mu_-$ is relevant and leads to a single Ising transition with $c=\frac{1}{2}$. This coupling translates to $\cos{\t_+} \cos{\t_-} = \frac{1}{2} [\cos{(2\t_\up)} + \cos{(2\t_\down)}]$, which breaks translation invariance explicitly. The central charge $c=\frac{1}{2}$ when translation symmetry is broken matches the analysis of Section~\ref{subsec:citocc}.

\subsubsection{Transition between $z$FM-I and SPT-I}Transition 2 in Fig.~\ref{fig:translation} connects the $z$FM-I phase (chargon condensate, spinon double condensate) and the SPT-I phase (chargon condensate, spinon insulator). Both phases can be accessed by perturbing the gapless parent Hamiltonian of Eq.~\eqref{eqn:LL1spin}, and we adopt the same bosonized description. Specifically, the transition between $z$FM-I and SPT-I phases is captured by 
\begin{align}
    \m{H}^{(2)}_{\text{transition}} &= {\cal H}_\text{LL}+\d \m{H}_{z\text{FM-I}} + \d \m{H}_{\text{SPT-I}}.
\end{align}
with $\d \m{H}_{\text{SPT-I}}$ as before in Eq.~(\ref{eqn:sg})
\begin{align}
    \d \m{H}_{z\text{FM-I}} &= - J_z (\sin{2\varphi_\up} + \sin{2\varphi_\down}). \label{eqn:zfmi}
\end{align}
Unlike the previous case, the interaction does not decouple under any canonical rotation of $\varphi,\theta$. However, we note that it is exactly self-dual for $K=1/\sqrt{2}$ and $J_{\text{SPT-I}} = J_z$. The self-duality here refers to the invariance of $\m{H}^{(2)}_{\text{transition}}$ under $J_z \leftrightarrow J_{\text{SPT-I}}, \varphi_{\up,\down} \leftrightarrow \t_\pm, \t_{\up,\down} \leftrightarrow \varphi_\pm$, which pinpoints the phase transition.

To determine the critical properties, we first drop all inter-wire couplings, which disallows  $\delta H_\text{SPT-I}$ and leaves $\delta H_\text{SPT-I}'\sim \sum_\l \cos 4 \theta_\l$ as the most relevant term pinning $\theta$. It results in two decoupled AT models~\cite{LecheminantGogolinNersesyan2002selfdual,JiangMotrunich2019DQCP} for a total central charge $c=1+1=2$. Each AT model is describable as a pair of Ising models coupled via marginal energy-energy interactions, i.e., 
\begin{align}
{\cal H}^\text{AT}_\l \sim  {\cal H}^\text{Ising}_{\l,A}+{\cal H}^\text{Ising}_{\l,B} + J \epsilon_{\l,A} \epsilon_{\l,B}.
\end{align}
The direct coupling between the Ising disorder operators $\mu_{\l,A}$ and $\mu_{\l,B}$, which would be relevant, is forbidden by translation symmetry. However, the product $\mu_{\uparrow,A}\mu_{\uparrow,B}\mu_{\downarrow,A}\mu_{\downarrow,B}$ is allowed and corresponds precisely to the sine-Gordon term of $\delta H_\text{SPT-I}$. It slaves one Ising variable to the other three, and we therefore expect a central charge $c=\frac{3}{2}$. Explicit breaking of translation symmetry further reduces the central charge to $c=1$, as we found in the lattice analysis of Section~\ref{subsec:spinontransition}.

The remaining transitions reduce to the ones analyzed above after adopting the dual spin representation of Section~\ref{sec:partontranslation}. Specifically, transition 3 is equivalent to transition 1 with $c=1$ ($c=\frac{1}{2}$ for broken translation symmetry) and transition 4 is equivalent to transition 2 with $c=\frac{3}{2}$ ($c=1$). The explicit analysis of both transitions using bosonized variables is presented in Appendix~\ref{appendix:transition}.

\section{Discussion}
\label{sec:conclusion}

We have constructed and analyzed spin ladders with discrete onsite symmetries that emulate  2D spinful bosons at half-filling. Specifically, we introduced an exact mapping of local spins to three partons (one chargon and two spinons), analogous to those used for accessing exotic states and transitions in 2D. Using this mapping, we obtained exactly solvable models for all relevant phases, where the partons form insulators, condensates, or SPT states. 
We further discussed phase transitions between these phases, in particular, critical points between translation-breaking and onsite-symmetry-breaking (but translation-invariant) phases.

Our 1D spin ladders emulate 2D spinful bosons at half-filling in two key respects: (i) by exhibiting analogous charge and spin symmetries and LSM constraints and (ii) by exhibiting exotic (deconfined) quantum phase transitions. We derived the field theories of those phase transitions and analyzed in detail how conventional criticality is recovered when translation symmetry is broken at the Hamiltonian level.

These findings highlight the broad extent of the close relationship between U(1) symmetric systems in 2D and $\mathbb{Z}_2$ symmetric systems in 1D. Further insights could be gained by pursuing these connections in several additional contexts.

First, it would be interesting to utilize Majorana fermion ladders with specific discrete symmetries to emulate phases and transitions of the 2D fermion models. In particular, we expect that Majorana ladders could capture many aspects of phase transitions between Chern insulators 
and charge density wave states in 2D electron systems~\cite{PanWuDasSarma2020moire}.  Specifically, the U(1)$_{\up,\down}$ symmetries of spinful electrons could be emulated by the $\mathbb{Z}_2$ fermion number parities $P_{\up,\down} = (-1)^{N_{\up,\down}}$ on the two legs $\uparrow,\downarrow$ of a 1D Majorana ladder. Moreover, a particle-hole symmetry of electrons at half-filling, $\m{C}^e: n^e_{\up,\down} \to 1- n^e_{\up,\down}$, could be mimicked by an onsite operator $\m{C} = i \gamma_\up \gamma_\down$, where $\gamma_{\uparrow,\downarrow}$ are Majorana fermions on the respective leg. Notably, $\m{C}$ anticommutes with $P_{\up,\down}$ within each unit cell, which leads to an LSM constraint. Similar to the symmetry $g_z$ studied here, the symmetry $\m{C}$ can thus emulate the half-filling of a 2D electron system. Notice, however, that Majorana ladders cannot capture metallic states in 2D fermion systems. 

Second, it would be appealing to study spin chains that emulate quantum phases and transitions of 2D systems with exotic symmetries, i.e., spatially modulated and non-invertible symmetries~\cite{SeibergShao2021fracton,McGreevy2023symmetry,Shao2023noninvertible}. For example, Ref.~\cite{LakeHermeleSenthil2022dbhm} investigated the phase diagram of 2D dipolar Bose-Hubbard models with dipole symmetries, revealing transitions beyond the conventional superfluid-insulator transition in the Bose-Hubbard model~\cite{Fisher1989SIT}. We expect close connections between these 2D systems and spin chains with multipole and non-invertible symmetries, such at those studied in Refs.~\onlinecite{LakeLeeHanSenthil2023dipole,SeifnashriShao2024noninvertiblespt}.

Finally, it would be desirable to systematically construct 1D models with specific symmetries that emulate 2D systems. One possible strategy is dimensional reduction. For example, Ref.~\cite{KarchTongTurner2019dualityweb2d} started with the U(1) Chern-Simons theory as the topological action for a massive (2+1)d Dirac fermion when the fermion mass is taken to infinity. Subsequently, dimensional reduction into (1+1)d led to an Arf-invariant of the $\mathbb{Z}_2$ gauge field (spin structure), which served as a topological action for a Majorana with infinite mass. The same topological response (Arf-invariant) arises in a lattice model of Majorana fermions known as the Kitaev chain~\cite{Kitaev2001majorana}. More generally, we expect dimensional reduction of a topological response to continuous symmetries in 2D to yield a specific response to discrete symmetries in 1D. This response could, in turn, guide the search for a suitable 1D lattice model to emulate the 2D physics.

\section*{Acknowledgement}
This work was supported by the Israel Science Foundation (ISF) under grant 2572/21. BH acknowledges the hospitality of the Institute
for Theoretical Physics in Cologne, supported by the
CRC 183 (funded by the DFG No. 277101999), where part of this work was carried out, and support by a Koshland fellowship from Weizmann Institute of Science.

\appendix

\section{Lieb-Schultz-Mattis theorem in our spin ladders}
\label{appendix:LSM}
In this section, we analyze the LSM theorem for our spin ladder using twisted boundary conditions. Given the symmetries in Eq.~(\ref{eqn:Z2sa}) and (\ref{eqn:Z2sb}), we twist the $g_z$ symmetry by modifying the translation operator as
\begin{align}
    \tilde{T}_x &\equiv% g^z_{r = 1} T_x =
    \s^z_{\uparrow,1} \s^z_{\downarrow,1} T_x.
\end{align}
Consequently, we have
\begin{align}
    \tilde{T}^N_x &= g_z.
\end{align}
Here $N$ is the total number of lattice sites, and we used $T^N_x = 1$. It is straightforward to verify that
\begin{align}
    g^{\l}_x \tilde{T}_x &= - \tilde{T}_x g^\l_x, \qquad \l = \uparrow,\downarrow.
\end{align}
This projective representation of the symmetry algebra signals an anomaly and excludes a symmetric gapped ground state~\cite{YaoOshikawa2021LSM}. 

\section{Review of the Kramers-Wannier duality for an Ising chain}
In this appendix, we review the well-known Kramers-Wannier duality for a transverse-field Ising model. In particular, we show how two applications of the duality, each introducing a $\mathbb{Z}_2$ gauge field, recover the original Ising model. This exercise introduces the conventions and strategies used to derive the results in the main text, in particular regarding emergent gauge fields. 

We start from the Hamiltonian 
\begin{align}
    H &= - \sum_r \s^z_r \s^z_{r+1} - h_x \sum_r \s^x_r - h_z \sum_r \s^z_r. \label{eqn:H0}
\end{align}
Under the Kramers-Wannier duality transformation
\begin{align}
\begin{split}
    \mu^x_{r+1/2} &= \s^z_r \s^z_{r+1}, \ \\ \mu^z_{r-1/2} \a^z_r \mu^z_{r+1/2} &= \s^x_r, \\ \ \a^x_r &= \s^z_r, 
    \end{split}\label{eqn:KW1}  
\end{align}
the Hamiltonian becomes
\begin{align}
    H &= -\sum_r \mu^x_{r+1/2} - h_x \sum_r \mu^z_{r-1/2} \a^z_r \mu^z_{r+1/2} - h_z \sum_r \a^x_r. \label{eqn:H1}
\end{align}
It includes the $\mathbb{Z}_2$ gauge field $\a_r$, which is subject to the gauge constraint
\begin{align}
 \a^x_r \a^x_{r+1} &= \mu^x_{r+1/2}, 
\end{align}
    
A second Kramers-Wannier duality transformation
\begin{equation}
\begin{split}
    \tau^x_r& = \mu^z_{r-1/2} \mu^z_{r+1/2},  \\ \ \tau^z_r \beta^z_{r+1/2} \tau^z_{r+1}& = \mu^x_{r+1/2}  \\ 
    \beta^x_{r+1/2} &= \mu^z_{r+1/2}, , \label{eqn:KW2}
      \end{split}
\end{equation}
with the constraint $\beta^x_{r-1/2} \beta^x_{r+1/2} = \tau^x_r$
leads to 
\begin{align}
    H &= -\sum_r \tau^z_r \beta^z_{r+1/2} \tau^z_{r+1} - h_x \sum_r \tau^x_r \a^z_r - h_z \sum_r \a^x_r, \label{eqn:H2}
\end{align}
where $\beta^{x/z}_{r+1/2}$ is the second $\mathbb{Z}_2$ gauge field. Eq.~(\ref{eqn:H0}) and (\ref{eqn:H2}) should be exactly the same, as $(\text{KW})^2 = 1$. To match the Hamiltonian in the two representations, we relabel the variables as
\begin{equation}
\begin{split}
    \tau'^x_r &= \tau^x_r \a^z_r, \ \tau'^z_r = \tau^z_r, \\
    \a'^x_r &= \a^x_r \tau^z_r , \ \a'^z_r = \a^z_r.
\end{split}
\end{equation}
In the new representation $\tau'$, we have
\begin{align}
    H &= - \sum_r \tau'^z_r \beta^z_{r+1/2} \tau'^z_{r+1} - h_x \sum_r \tau'^x_r - h_z \sum_r \a'^x_r \tau'^z_r.
\end{align}
The gauge constraints in Eq.~(\ref{eqn:KW1}) and (\ref{eqn:KW2}) become
\begin{align}
    \a'^x_r \tau'^z_r \a'^x_{r+1} \tau'^z_{r+1} &= \tau^z_r \beta^z_{r+1/2} \tau^z_{r+1} = \tau'^z_r \beta^z_{r+1/2} \tau'^z_{r+1}, \nonumber \\
    \Longrightarrow \a'^x_r \a'^x_{r+1} &= \beta^z_{r+1/2}. \label{eqn:Gauss1}
\end{align}
and
\begin{align}
    \beta^x_{r-1/2} \beta^x_{r+1/2} &= \tau'^x_r \a'^z_r, \label{eqn:Gauss2}
\end{align}
respectively. 
A second relabeling of variables
\begin{equation}
\begin{split}
    \beta'^z_{r+1/2} &= \beta^z_{r+1/2} \a'^x_r \a'^x_{r+1}, \\ \beta'^x_{r-1/2} \beta'^x_{r+1/2} &= \tau'^x_r \a''^z_r, \\
    \tau''^z_r &= \tau'^z_r \a'^x_r, \\ \tau''^x_r &= \tau'^x_r, \\
    \a''^z_r &= \beta^x_{r-1/2} \a'^z_r \tau'^x_r \beta^x_{r+1/2}, \\ \a''^x_r &= \a'^x_r
\end{split}
\end{equation} 
leads to
\begin{align}
    H &= -\sum_r \tau''^z_r \beta'^z_{r+1/2} \tau''^z_{r+1} - h_x \sum_r \tau''^x_r - h_z \sum_r \tau''^z_r. \label{eqn:H4}
\end{align}
The gauge constraints are
\begin{align}
    \beta'^z_{r+1/2} &= 1, \ \a''^z_r =1,
\end{align}
recovering Eq.~(\ref{eqn:H0}). 

\section{Evaluation of partition functions in Section~\ref{subsec:VBS}}
\label{app:VBS}

In this appendix, we demonstrate the calculation of partition functions for the VBS states discussed in Section~\ref{subsec:VBS}, using spin ladders defined on a circle with circumference $2N$. For illustration, we explain how to compute the partition functions for the state in Fig.~\ref{fig:VBSSPT}(b). A Hamiltonian with periodic boundary conditions that has this state as the ground state is
\begin{align}
    H &= -\sum_{\l = \up, \down} \sum^N_{r=1} (\s^x_{\l, 2r} \s^x_{\l,2r+1} + \s^z_{\l,2r} \s^z_{\l,2r+1}).
\end{align}
 We impose twisted boundary condition for the flux configuration $h=(h^\uparrow_x,h^\downarrow_x,h_z)$ by modifying the terms crossing the $(2N,1)$ link as

\begin{equation}
\begin{split}
    H_{\text{bdy}} &= -h_z (\s^x_{\up,2N} \s^x_{\up,1} + \s^x_{\down,2N} \s^x_{\down,1})  \\
    &\quad - h^\up_x \s^z_{\up,2N} \s^z_{\up,1} - h^\down_x \s^z_{\down,2N} \s^z_{\down,1},
    \end{split}
\end{equation}
and other terms in $H$ are unchanged. At zero temperature, only ground states contribute to the partition function. Thus we have
\begin{align}
    Z[g,h] &= {}_h \langle \text{GS}| g |\text{GS}\rangle_h,
\end{align}
where $|\text{GS}\rangle_h$ represents the ground state of the Hamiltonian $H[h]$. This calculation yields the results in the third column of Table~\ref{tab:pf}.

\begin{widetext}
\section{Derivations of the parton representation of local spins}
\label{app:dictionary}
In this appendix, we discuss the details of the Kramers-Wannier dualities from the local spins $\s$ to the domain-walls $\mu$ and finally to the partons $\tau$. We start with two decoupled Ising chains. 
The local degrees of freedom are $\s^{x/z}_{\l,r}$. We perform the first duality transformation
\begin{align}
 \s^z_{\l,r} \s^z_{\l,r+1} &= \mu^x_{\l,r+1/2},  \ &&\s^x_{\l,r} = \mu^z_{\l,r-1/2} \a^z_{\l,r} \mu^z_{\l,r+1/2}, \nonumber \\
    \a^x_{\l,r} &= \s^z_{\l,r}, \ 
    &&\text{gauge constraint: } \a^x_{\l,j} \a^x_{\l,r+1} = \mu^x_{\l,r+1/2}.
\end{align}
Here, $\a^{x/z}_{\l,r}$ are the two $\mathbb{Z}_2$ gauge fields, which are related to $\a_{c,n}$ used in the main text as $\a_{c,r} = \a_{\up,r} \a_{\down,r}$ and $\a_{n,r} = \a_{\down,r}$. Now we introduce separate variables for domain walls on even add odd links according to
\begin{equation}
\begin{split}
    \mu^{x/z}_{\l,2r+1/2} &= \mu^{x/z}_{\l,+,r+1/2} , \qquad \mu^{x/z}_{\l,2r-1/2} = \mu^{x/z}_{\l,-,r+1/2}.
    \end{split}
\end{equation}
Subsequently, we perform a charge-spin decomposition, which is a relabeling of domain-walls. 
\begin{align}
    &\mu^z_{c,2r+1/2} = \mu^z_{\uparrow,2r+1/2} \mu^z_{\downarrow,2r+1/2}, \qquad&&
    \mu^x_{c,2r+1/2} = \mu^x_{\uparrow,2r+1/2},
     \\ 
   & \mu^z_{c,2r-1/2} = \mu^z_{\uparrow,2r-1/2} \mu^z_{\downarrow,2r-1/2}, \qquad&&
    \mu^x_{c,2r-1/2} = \mu^x_{\up, 2r-1/2},
     \\
   & \mu^z_{n,2r+1/2} = \mu^z_{\downarrow,2r+1/2}, \qquad&&
    \mu^x_{n,2r+1/2} = \mu^x_{\uparrow, 2r+1/2} \mu^x_{\downarrow,2r+1/2},
    \\
   & \mu^z_{n,2r-1/2} = \mu^z_{\uparrow,2r-1/2}, \qquad&&
    \mu^x_{n,2r-1/2} = \mu^x_{\uparrow,2r-1/2} \mu^x_{\downarrow, 2r-1/2}.
    \end{align}
To form gauge invariant operators in the charge/neutral basis, we note that $\mu^x_{a,r+1/2}$ $(a = c,n)$ are always gauge invariant, while $\mu^z_{a,r+1/2}$, requires $\a^z_{\l,r}$ for gauge invariance. The gauge invariant operators and their microscopic representations are
\begin{align}
    \mu^x_{c,2r+1/2} &= \s^z_{\uparrow,2r} \s^z_{\uparrow,2r+1}, \ &&\mu^z_{c,2r-1/2} \a^z_{\uparrow,2r} \a^z_{\downarrow,2r} \mu^z_{c,2r+1/2} = \s^x_{\uparrow,2r} \s^x_{\downarrow,2r}, \\
    \mu^x_{c,2r-1/2} &= \s^z_{\uparrow,2r-1} \s^z_{\uparrow,2r} , \ 
     &&\mu^z_{c,2r+1/2} \a^z_{\uparrow,2r+1} \a^z_{\downarrow,2r+1} \mu^z_{c,2r+3/2} = \s^x_{\uparrow,2r+1} \s^x_{\downarrow,2r+1}, \\
    \mu^x_{n,2r+1/2} &= \s^z_{\uparrow,2r} \s^z_{\uparrow,2r+1} \s^z_{\downarrow,2r} \s^z_{\downarrow,2r+1} , \ &&\mu^z_{n,2r-1/2} \a^z_{\uparrow,2r} \a^z_{\uparrow,2r+1} \mu^z_{n,2r+3/2} = \s^x_{\uparrow,2r} \s^x_{\uparrow,2r+1}, \\
    \mu^x_{n,2r-1/2} &= \s^z_{\uparrow,2r-1} \s^z_{\uparrow,2r} \s^z_{\downarrow,2r-1} \s^z_{\downarrow,2r}, \  &&\mu^z_{n,2r+1/2} \a^z_{\downarrow,2r+1} \a^z_{\downarrow,2r+2} \mu^z_{n,2r+5/2} = \s^x_{\downarrow,2r+1} \s^x_{\downarrow,2r+2}. \label{eqn:cndecouple}
\end{align}
In the charge/neutral basis, the Gauss' constraints become
\begin{equation}
    \begin{split}
    G_{\uparrow,2r+1/2} &= \a^x_{\uparrow,2r} \mu^x_{c,2r+1/2} \a^x_{\uparrow,2r+1} = 1, \\
    G_{\downarrow,2r+1/2} &= \a^x_{\downarrow,2r} \mu^x_{c,2r+1/2} \mu^x_{n,2r+1/2} \a^x_{\downarrow,2r+1} = 1, \\
    G_{\uparrow,2r-1/2} &= \a^x_{\uparrow,2r-1} \mu^x_{c,2r-1/2} \mu^x_{n,2r-1/2} \a^x_{\uparrow,2r} = 1, \\
    G_{\downarrow,2r-1/2} &= \a^x_{\downarrow,2r-1} \mu^x_{c,2r-1/2} \a^x_{\downarrow,2r} = 1.
    \end{split}. \label{eqn:Gaussmu}
\end{equation}
From the second and third lines of Eq.~(\ref{eqn:cndecouple}), we can see that both charge sectors couple to $\a^z_c$. Consequently,  we can group them into a single charge sector $\mu^z_{c,r+1/2}$, irrespective of the even/odd sector. The same is not possible for the neutral even/odd sectors because they couple to $\a_c \a_n$ and $\a_n$, respectively. 

The manipulation above left us with three domain-wall operators $\mu^{x/z}_{c,r+1/2}, \mu^{x/z}_{n,2r+1/2}, \mu^{x/z}_{n,2r-1/2}$. We perform a separate duality transformations on each of them according to
\begin{align}
    \mu^z_{c,r-1/2} \mu^z_{c,r+1/2} &= \tau^x_{c,r}, \  &&\mu^x_{c,r+1/2} = \tau^z_{c,r} \beta^z_{c,r+1/2} \tau^z_{c,r+1}, \nonumber  \\ \beta^x_{c,r+1/2} &= \mu^z_{c,r+1/2}, \
    &&\text{gauge constraint: } \beta^x_{c,r-1/2} \beta^x_{c,r+1/2} = \tau^x_{c,r}, \nonumber
     \\ 
    \mu^z_{n,2r+1/2} \mu^z_{n,2r+5/2} &= \tau^x_{n,2r+2}, \  &&\mu^x_{n,2r+1/2} = \tau^z_{n,2r} \beta^z_{n,2r+1/2} \tau^z_{n,2r+2},  \label{eqn:KWII}  \\ \beta^x_{n,2r+1/2} &= \mu^z_{n,2r+1/2}, \
    &&\text{gauge constraint: }  \beta^x_{n,2r+1/2} \beta^x_{n,2r+5/2} = \tau^x_{n,2r+2}, \nonumber
     \\
    \mu^z_{n,2r-1/2} \mu^z_{n,2r+3/2} &= \tau^x_{n,2r+1}, \ &&\mu^x_{n,2r-1/2} = \tau^z_{n,2r-1} \beta^z_{n,2r-1/2} \tau^z_{n,2r+1}, \nonumber \\ \beta^x_{n,2r-1/2} &= \mu^z_{n,2r-1/2}, \
    &&\text{gauge constraint: } \beta^x_{n,2r-1/2} \beta^x_{n,2r+3/2} = \tau^x_{n,2r+1}. \nonumber  
\end{align}
After the duality transformation, the Gauss' constraints in Eq.~(\ref{eqn:Gaussmu}) become
\begin{equation}
    \begin{split}
    G_{\uparrow,2r+1/2} &= \a^x_{\uparrow,2r} \a^x_{\uparrow,2r+1} \tau^z_{c,2r} \beta^z_{c,2r+1/2} \tau^z_{c,2r+1} = 1, \\
    G_{\downarrow,2r+1/2} &= \a^x_{\downarrow,2r} \a^x_{\downarrow,2r+1} \tau^z_{c,2r} \beta^z_{c,2r+1/2} \tau^z_{c,2r+1} \tau^z_{n,2r} \beta^z_{n,2r+1/2} \tau^z_{n,2r+2} = 1, \\
    G_{\uparrow,2r-1/2} &= \a^x_{\uparrow,2r-1} \a^x_{\uparrow,2r} \tau^z_{c,2r-1} \beta^z_{c,2r-1/2} \tau^z_{c,2r} \tau^z_{n,2r-1} \beta^z_{n,2r-1/2} \tau^z_{n,2r+1} = 1, \\
    G_{\downarrow,2r-1/2} &= \a^x_{\downarrow,2r-1} \a^x_{\downarrow,2r} \tau^z_{c,2r-1} \beta^z_{c,2r-1/2} \tau^z_{c,2r}=1.
    \end{split} \label{eqn:Gausstau}
\end{equation}

We simplify the Hamiltonian by integrating out some gauge fields. For this purpose we eliminate $\a$ gauge fields in Eq.~(\ref{eqn:Gausstau}) by redefining variables as
\begin{align}
    \a'^x_{\uparrow,2r} &= \a^x_{\uparrow,2r} \tau^z_{c,2r} \tau^z_{n,2r+1}, \ &&\tau'^x_{c,r} = \tau^x_{c,r} \a^z_{\uparrow,r} \a^z_{\downarrow,r}, \nonumber \\
    \a'^x_{\uparrow,2r+1} &= \a^x_{\uparrow,2r+1} \tau^z_{c,2r+1} \tau^z_{n,2r+1}, \ &&\tau'^x_{n,2r} = \tau^x_{n,2r} \a^z_{\downarrow,2r-1} \a^z_{\downarrow,2r}, \nonumber \\
    \a'^x_{\downarrow,2r} &= \a^x_{\downarrow,2r} \tau^z_{c,2r} \tau^z_{n,2r}, \ &&\tau'^x_{n,2r+1} = \tau^x_{n,2r+1} \a^z_{\uparrow,2r} \a^z_{\uparrow,2r+1}, \nonumber \\
    \a'^x_{\downarrow,2r+1} &= \a^x_{\downarrow,2r+1} \tau^z_{c,2r+1} \tau^z_{n,2r+2},
    \ &&\tau'^z_{c,r} = \tau^z_{c,r}, \\
    \a'^z_{\uparrow,2r} &= \a^z_{\uparrow,2r}, \ &&\tau'^z_{n,2r} = \tau^z_{n,2r}, \nonumber \nonumber \\
    \a'^z_{\uparrow,2r+1} &= \a^z_{\uparrow,2r+1}, \ &&\tau'^z_{n,2r+1} = \tau^z_{n,2r+1}, \nonumber \\
    \a'^z_{\downarrow,2r} &= \a^z_{\downarrow,2r}, \nonumber \\
    \a'^z_{\downarrow,2r+1} &= \a^z_{\downarrow,2r+1}. \nonumber
\end{align}
Consequently, Eq.~(\ref{eqn:Gausstau}) becomes
\begin{equation}
    \begin{split}
       G_{\uparrow,2r+1/2} &= \a'^x_{\uparrow,2r} \beta^z_{c,2r+1/2} \a'^x_{\uparrow,2r+1} = 1, \\
       G_{\downarrow,2r+1/2} &= \a'^x_{\downarrow,2r} \beta^z_{c,2r+1/2} \beta^z_{n,2r+1/2} \a'^x_{\downarrow,2r+1} = 1, \\
       G_{\uparrow,2r-1/2} &= \a'^x_{\uparrow,2r-1} \beta^z_{c,2r-1/2} \beta^z_{n,2r-1/2} \a'^x_{\uparrow,2r} =1, \\
       G_{\downarrow,2r-1/2} &= \a'^x_{\downarrow,2r-1} \beta^z_{c,2r-1/2} \a'^x_{\downarrow,2r} = 1.
    \end{split}. \label{eqn:GaussIII}
\end{equation}
Together with the three constraints in Eq.~(\ref{eqn:KWII}), we have
\begin{equation}
    \begin{split}
    G_{c,r} &= \beta^x_{c,r-1/2} \beta^x_{c,r+1/2} \tau'^x_{c,r} \a'^z_{\uparrow,r} \a'^z_{\downarrow,r} = 1, \\
    G_{n,2r+2} &= \beta^x_{n,2r+1/2} \beta^x_{n,2r+5/2} \tau'^x_{n,2r+2} \a'^z_{\downarrow,2r+1} \a'^z_{\downarrow,2r+2} =1, \\
    G_{n,2r+1} &= \beta^x_{n,2r-1/2} \beta^x_{n,2r+3/2} \tau'^x_{n,2r+1} \a'^z_{\uparrow,2r} \a'^z_{\uparrow,2r+1}=1,
    \end{split}\label{eqn:GaussII}
\end{equation}
which leads to 
\begin{align}
    \tau'^x_{c,2r+1} \tau'^x_{c,2r+2} \tau'^x_{n,2r+1} \tau'^x_{n,2r+2} &= \beta^x_{c,2r+1/2} \beta^x_{c,2r+5/2} \beta^x_{n,2r-1/2} \beta^x_{n,2r+1/2} \beta^x_{n,2r+3/2} \beta^x_{n,2r+5/2} \a'^z_{\uparrow,2r} \a'^z_{\uparrow,2r+2}, \label{eqn:CNconstraint}
\end{align}
by taking the products of $G_{c,2r+1} G_{c,2r+2} G_{n,2r+1} G_{n,2r+2} = 1$. The above equation can be further simplified by changing the basis of the gauge fields (relabeling of gauge fields) as %\david{You don't 'introduce a gauge field' I suppose. you are doing a change of basis, clarify} a new gauge field $\g_r$:
\begin{align}
    \g^x_{2r+1/2} = \beta^x_{c,2r+1/2} \beta^x_{n,2r-1/2} \beta^x_{n,2r+1/2} \a'^z_{\uparrow,2r}, \ \ 
    \g^z_{2r+1/2} = \beta^z_{n,2r+1/2},
\end{align}
such that Eq.~(\ref{eqn:CNconstraint}) becomes
\begin{align}
    \tau'^x_{c,2r+1} \tau'^x_{c,2r+2} \tau'^x_{n,2r+1} \tau'^x_{n,2r+2} &= \g^x_{2r+1/2} \g^x_{2r+5/2}.
\end{align}
From Eq.~(\ref{eqn:GaussIII}), we have
\begin{equation}
    \begin{split}
    G_{\uparrow,2r+1/2} G_{\downarrow,2r+1/2} =1 &\Rightarrow \a'^x_{\uparrow,2r} \a'^x_{\downarrow,2r} \a'^x_{\uparrow,2r+1} \a'^x_{\downarrow,2r+1} = \beta^z_{n,2r+1/2}, \\
    G_{\uparrow,2r+1/2} G_{\uparrow,2r-1/2} = 1 &\Rightarrow \a'^x_{\uparrow,2r-1} \a'^x_{\uparrow,2r+1} = \beta^z_{c,2r-1/2} \beta^z_{n,2r-1/2} \beta^z_{c,2r+1/2}, \\
    G_{\downarrow,2r+1/2} G_{\downarrow,2r-1/2} = 1 &\Rightarrow \a'^x_{\downarrow,2r-1} \a'^x_{\downarrow,2r+1} = \beta^z_{c,2r-1/2} \beta^z_{c,2r+1/2} \beta^z_{n,2r+1/2}.
    \end{split}
\end{equation}
We relabel gauge fields as  
\begin{align}
    &\g^{(1)x}_{2r-1/2} = \beta^x_{n,2r-1/2}, \quad
    &&\g^{(2)x}_{2r+1/2} = \beta^x_{c,2r+1/2}, \quad
     &&\g^{(3)x}_{2r-1/2} = \beta^x_{c,2r-1/2}, \\
    &\g^{(4)x}_{2r} = \a'^x_{\uparrow,2r} \a'^x_{\downarrow,2r} \beta^z_{n,2r+1/2}, \quad
    &&\g^{(5)x}_{2r+1} = \a'^x_{\uparrow,2r+1} \a'^x_{\downarrow,2r+1}, \quad
     &&\g^{(6)x}_r = \a'^x_{\downarrow,r}, \\ 
    &\g^{(1)z}_{2r-1/2} = \beta^z_{n,2r-1/2} \beta^z_{n,2r+1/2}, \quad
    &&\g^{(2)z}_{2r+1/2} = \beta^z_{c,2r+1/2} \beta^z_{n,2r+1/2}, \quad
     &&\g^{(3)z}_{2r-1/2} = \beta^z_{c,2r-1/2}, \\
    &\g^{(4)z}_{2r} = \a'^z_{\uparrow,2r}, \quad
    &&\g^{(5)z}_{2r+1}= \a'^z_{\uparrow,2r+1}, \quad
   && \g^{(6)z}_r = \a'^z_{\uparrow,r} \a'^z_{\downarrow,r}.
    \end{align}
The constraints in Eq.~(\ref{eqn:GaussII}) become
\begin{equation}
    \begin{split}
    G_{c,2r} &= \g^{(2)x}_{2r+1/2} \g^{(3)x}_{2r-1/2} \tau'^x_{c,2r} \g^{(6)z}_{2r} = 1, \\
    G_{c,2r+1} &= \g^{(2)x}_{2r+1/2} \g^{(3)x}_{2r+3/2} \tau'^x_{c,2r+1} \g^{(6)z}_{2r+1} = 1, \\
    G_{n,2r+1} &= \g^{(1)x}_{2r-1/2} \g^{(1)x}_{2r+3/2} \tau'^x_{n,2r+1} \g^{(4)z}_{2r} \g^{(5)z}_{2r+1} = 1,
    \end{split} \label{eqn:GaussIV}
\end{equation}
and the constraints in Eq.~(\ref{eqn:GaussIII}) become
\begin{equation}
    \begin{split}
    G_{\uparrow,2r+1/2} &= \g^{(4)x}_{2r} \g^{(6)x}_{2r} \g^{(2)z}_{2r+1/2} \g^{(5)x}_{2r+1} \g^{(6)x}_{2r+1} = 1, \\
    G_{\downarrow,2r+1/2} &= \g^{(6)x}_{2r} \g^{(2)z}_{2r+1/2} \g^{(6)x}_{2r+1} = 1, \\
    G_{\uparrow,2r-1/2} &= \g^{(5)x}_{2r-1} \g^{(6)x}_{2r-1} \g^{(3)z}_{2r-1/2} \g^{(1)z}_{2r-1/2} \g^{(4)x}_{2r} \g^{(6)x}_{2r} = 1, \\
    G_{\downarrow,2r-1/2} &= \g^{(6)x}_{2r-1} \g^{(3)z}_{2r-1/2} \g^{(6)x}_{2r} = 1.
    \end{split} \label{eqn:Gaussgamma}
\end{equation}
Now we relabel the variables as
\begin{align}
 &\g'^{(1)x}_{2r-1/2} = \g^{(1)x}_{2r-1/2},  &&\qquad \g'^{(1)z}_{2r-1/2} = \g^{(1)z}_{2r-1/2} \g^{(5)x}_{2r-1} \g^{(5)x}_{2r+1}, \\
    &\g'^{(2)x}_{2r+1/2} = \g^{(2)x}_{2r+1/2} &&\qquad \g'^{(2)z}_{2r+1/2} = \g^{(2)z}_{2r+1/2} \g^{(6)x}_{2r} \g^{(6)x}_{2r+1}, \\
    &\g'^{(3)x}_{2r-1/2} = \g^{(3)x}_{2r-1/2} &&\qquad \g'^{(3)z}_{2r-1/2} = \g^{(3)z}_{2r-1/2} \g^{(6)x}_{2r-1} \g^{(6)x}_{2r}, \\
    &\g'^{(4)x}_{2r} = \g^{(4)x}_{2r} \g^{(5)x}_{2r+1}, &&\qquad  \g'^{(4)z}_{2r} = \g^{(4)z}_{2r}, \\
    &\g'^{(5)x}_{2r+1} = \g^{(5)x}_{2r+1} &&\qquad \g'^{(5)z}_{2r+1} = \g^{(5)z}_{2r+1} \g^{(4)z}_{2r} \g^{(1)x}_{2r-1/2} \g^{(1)x}_{2r+3/2} \tau'^x_{n,2r+1}, \\
    &\g'^{(6)x}_{2r} = \g^{(6)x}_{2r} &&\qquad \g'^{(6)z}_{2r} = \g^{(6)z}_{2r} \g^{(3)x}_{2r-1/2} \g^{(2)x}_{2r+1/2} \tau'^x_{c,2r}, \\
    &\g'^{(6)x}_{2r+1} = \g^{(6)x}_{2r+1} &&\qquad \g'^{(6)z}_{2r+1} = \g^{(6)z}_{2r+1} \g^{(2)x}_{2r+1/2} \g^{(3)x}_{2r+3/2} \tau'^x_{c,2r+1}, \\
    &\tau''^x_{c,r} = \tau'^x_{c,r} &&\qquad \tau''^z_{c,r} = \tau'^z_{c,r} \g^{(6)x}_r, \\
    &\tau''^x_{n,2r+1} = \tau'^x_{n,2r+1}, &&\qquad \tau''^z_{n,2r+1} = \tau'^z_{n,2r+1} \g^{(5)x}_{2r+1}.    
\end{align}

In the relabeled variables, the constraints in Eq.~(\ref{eqn:GaussIV}) and (\ref{eqn:Gaussgamma}) become 
\begin{equation}
\begin{split}
    &G_{\uparrow,2r+1/2} = \g'^{(4)x}_{2r} \g'^{(2)z}_{2r+1/2} = 1,  \\ 
    &G_{\downarrow,2r+1/2} = \g'^{(2)z}_{2r+1/2} = 1,  \\
    &G_{\uparrow,2r-1/2} = \g'^{(1)z}_{2r-1/2} \g'^{(4)x}_{2r} \g'^{(3)z}_{2r-1/2} = 1,  \\
    &G_{\downarrow,2r-1/2} = \g'^{(3)z}_{2r-1/2} = 1,
      \\
      &G_{c,2r+1} = \g'^{(6)z}_{2r+1} = 1,  \\
      &G_{c,2r} = \g'^{(6)z}_{2r} = 1,  \\
      &G_{n,2r+1} = \g'^{(5)z}_{2r+1} = 1. 
    \end{split}
\end{equation}

Then we have
\begin{equation}
\begin{split}
    &\g'^{(1)z}_{2r-1/2} =\g'^{(4)x}_{2r} =\g'^{(2)z}_{2r+1/2} =\g'^{(5)z}_{2r+1} = \g'^{(3)z}_{2r-1/2}  = 1,  
   \end{split}
\end{equation}
together with the Gauss constraint
\begin{align}
    \tau''^x_{c,2r+1} \tau''^x_{c,2r+2} \tau''^x_{n,2r+1} \tau''^x_{n,2r+2} &= \g^x_{2r+1/2} \g^x_{2r+5/2}.
\end{align}
Consequently, we have only a single gauge field $\g_{2r+1/2}$ left.

Here we express the local spin operators $\s_{\l,r}$ in terms of the parton operators $\tau''_{c/n,r}$ and the gauge fields $\g_{2r+1/2}$. 

\begin{flalign}
   \quad \s^z_{\uparrow,2r} \s^z_{\uparrow,2r+1} &= \mu^x_{\uparrow,2r+1/2} = \mu^x_{c,2r+1/2} = \tau^z_{c,2r} \beta^z_{c,2r+1/2} \tau^z_{c,2r+1} \nonumber &\\
    &= \tau'^z_{c,2r} \beta^z_{c,2r+1/2} \tau'^z_{c,2r+1} = \tau'^z_{c,2r} \g^{(2)z}_{2r+1/2} \g^z_{2r+1/2} \tau'^z_{c,2r+1} \nonumber &\\
    &= (\tau''^z_{c,2r} \g^{(6)x}_{2r}) (\g'^{(2)z}_{2r+1/2} \g^{(6)x}_{2r} \g^{(6)x}_{2r+1}) \g^z_{2r+1/2} (\tau''^z_{c,2r+1} \g^{(6)x}_{2r+1}) \nonumber &\\
    &= \tau''^z_{c,2r} \g^z_{2r+1/2} \tau''^z_{c,2r+1},&
    \end{flalign}
    \begin{flalign}
   \quad  \s^z_{\downarrow,2r-1} \s^z_{\downarrow,2r} &= \mu^x_{\downarrow,2r-1/2} = \mu^x_{c,2r-1/2} = \tau^z_{c,2r-1} \beta^z_{c,2r-1/2} \tau^z_{c,2r} \nonumber &\\
    &= \tau'^z_{c,2r-1} \beta^z_{c,2r-1/2} \tau'^z_{c,2r} = \tau'^z_{c,2r-1} \g^{(3)z}_{2r-1/2} \tau'^z_{c,2r} \nonumber &\\
    &= (\tau''^z_{c,2r-1} \g^{(6)x}_{2r-1}) (\g'^{(3)z}_{2r-1/2} \g^{(6)x}_{2r-1} \g'^{(6)x}_{2r}) (\tau''^z_{c,2r} \g^{(6)x}_{2r}) \nonumber &\\
    &= \tau''^z_{c,2r-1} \tau''^z_{c,2r},&
    \end{flalign}
    \begin{flalign}
    \quad \s^z_{\uparrow,2r-1} \s^z_{\uparrow,2r} &= \mu^x_{\uparrow,2r-1/2} = \mu^x_{n,2r-1/2} \mu^x_{c,2r-1/2} \nonumber &\\
    &= \tau^z_{n,2r-1} \beta^z_{n,2r-1/2} \tau^z_{n,2r+1} \cdot \tau^z_{c,2r-1} \beta^z_{c,2r-1/2} \tau^z_{c,2r} \nonumber &\\
    &= \tau'^z_{n,2r-1} \g^{(1)z}_{2r-1/2} \g^z_{2r+1/2} \tau'^z_{n,2r+1} \cdot \tau'^z_{c,2r-1} \g^{(3)z}_{2r-1/2} \tau'^z_{c,2r} \nonumber &\\
    &= \tau''^z_{n,2r-1} (\g'^{(1)z}_{2r-1/2} \g^{(5)x}_{2r-1} \g^{(5)x}_{2r+1}) \g^z_{2r+1/2} (\tau''^z_{n,2r+1} \g^{(5)x}_{2r+1}) (\tau''^z_{c,2r-1} \g^{(6)x}) (\g'^{(3)z}_{2r-1/2} \g^{(6)x}_{2r-1} \g^{(6)x}_{2r}) (\tau''^z_{c,2r} \g^{(6)x}_{2r}) \nonumber &\\
    &= \tau''^z_{n,2r-1} \g'^{(1)z}_{2r-1/2} \g^{(5)x}_{2r-1} \g^z_{2r+1/2} \tau''^z_{n,2r+1} \tau''^z_{c,2r-1} \g'^{(3)z}_{2r-1/2} \tau''^z_{c,2r} \nonumber &\\
    &= \tau''^z_{n,2r-1} \g^z_{2r+1/2} \tau''^z_{n,2r+1} \tau''^z_{c,2r-1} \tau''^z_{c,2r}, &
    \end{flalign}
    \begin{flalign}
    \quad \s^z_{\downarrow,2r} \s^z_{\downarrow,2r+1} &= \mu^x_{\downarrow,2r+1/2} = \mu^x_{c,2r+1/2} \mu^x_{n,2r+1/2} \nonumber &\\
    &= \tau^z_{c,2r} \beta^z_{c,2r+1/2} \tau^z_{c,2r+1} \cdot \tau^z_{n,2r} \beta^z_{n,2r+1/2} \tau^z_{n,2r+2} \nonumber &\\
    &= \tau'^z_{c,2r} \g^{(2)z}_{2r+1/2} \g^z_{2r+1/2} \tau'^z_{c,2r+1} \cdot \tau'^z_{n,2r} \g^z_{2r+1/2} \tau'^z_{n,2r+2} \nonumber &\\
    &= (\tau''^z_{c,2r} \g^{(6)x}_{2r}) (\g'^{(2)z}_{2r+1/2} \g^{(6)x}_{2r} \g^{(6)x}_{2r+1}) (\tau''^z_{c,2r+1} \g^{(6)x}_{2r+1}) (\tau''^z_{n,2r} \tau''^z_{n,2r+2}) \nonumber &\\
    &= \tau''^z_{c,2r} \tau''^z_{c,2r+1} \tau''^z_{n,2r} \tau''^z_{n,2r+2}. &
    \end{flalign}
\begin{flalign}
  \quad  \s^x_{\uparrow,2r} \s^x_{\downarrow,2r} &= \mu^z_{\uparrow,2r-1/2} \a^z_{\uparrow,2r} \mu^z_{\uparrow,2r+1/2} \cdot \mu^z_{\downarrow,2r-1/2} \a^z_{\downarrow,2r} \mu^z_{\downarrow,2r+1/2} \nonumber &\\
    &= \mu^z_{c,2r-1/2} \mu^z_{c,2r+1/2} \a^z_{\uparrow,2r} \a^z_{\downarrow,2r} = \tau^x_{c,2r} \a^z_{\uparrow,2r} \a^z_{\downarrow,2r} = \tau'^x_{c,2r} \nonumber &\\
    &= \tau''^x_{c,2r}, &
    \end{flalign}
    \begin{flalign}
   \quad \s^x_{\uparrow,2r+1} \s^x_{\downarrow,2r+1} &= \mu^z_{\uparrow,2r+1/2} \a^z_{\uparrow,2r+1} \mu^z_{\uparrow,2r+3/2} \cdot \mu^z_{\downarrow,2r+1/2} \a^z_{\downarrow,2r+1} \mu^z_{\downarrow,2r+3/2} \nonumber &\\
    &= \mu^z_{c,2r+1/2} \mu^z_{c,2r+3/2} \a^z_{\uparrow,2r+1} \a^z_{\downarrow,2r+1} = \tau^x_{c,2r+1} \a^z_{\uparrow,2r+1} \a^z_{\downarrow,2r+1} = \tau'^x_{c,2r+1} \nonumber &\\
    &= \tau''^x_{c,2r+1}, &
    \end{flalign}
    \begin{flalign}
   \quad \s^x_{\uparrow,2r} \s^x_{\uparrow,2r+1} &= \mu^z_{\uparrow,2r-1/2} \a^z_{\uparrow,2r} \mu^z_{\uparrow,2r+1/2} \cdot \mu^z_{\uparrow,2r+1/2} \a^z_{\uparrow,2r+1} \mu^z_{\uparrow,2r+3/2} = \mu^z_{n,2r-1/2} \a^z_{\uparrow,2r} \a^z_{\uparrow,2r+1} \mu^z_{n,2r+3/2} \nonumber &\\
    &= \tau^x_{n,2r+1} \a^z_{\uparrow,2r} \a^z_{\uparrow,2r+1} = \tau'^x_{n,2r+1} \nonumber &\\
    &= \tau''^x_{n,2r+1}, &
     \end{flalign}
    \begin{flalign}
   \quad \s^x_{\downarrow,2r-1} \s^x_{\downarrow,2r} &= \mu^z_{\downarrow,2r-3/2} \a^z_{\downarrow,2r-1} \mu^z_{\downarrow,2r-1/2} \cdot \mu^z_{\downarrow,2r-1/2} \a^z_{\downarrow,2r} \mu^z_{\downarrow,2r+1/2} \nonumber &\\
    &= \mu^z_{\downarrow,2r-3/2} \a^z_{\downarrow,2r-1} \a^z_{\downarrow,2r} \mu^z_{\downarrow,2r+1/2} \nonumber &\\
    &= \mu^z_{n,2r-3/2} \a^z_{\downarrow,2r-1} \a^z_{\downarrow,2r} \mu^z_{n,2r+1/2} \nonumber &\\
    &= \tau^x_{n,2r} \a^z_{\downarrow,2r-1} \a^z_{\downarrow,2r} = \tau'^x_{n,2r} \nonumber &\\
    &= \tau''^x_{n,2r}, &
     \end{flalign}
    \begin{flalign}
   \quad \s^x_{\downarrow,2r} \s^x_{\downarrow,2r+1} &= (\mu^z_{\downarrow,2r-1/2} \a^z_{\downarrow,2r} \mu^z_{\downarrow,2r+1/2}) (\mu^z_{\downarrow,2r+1/2} \a^z_{\downarrow,2r+1} \mu^z_{\downarrow,2r+3/2}) \nonumber &\\
    &= (\mu^z_{\downarrow,2r-1/2} \mu^z_{\downarrow,2r+3/2}) (\a^z_{\downarrow,2r} \a^z_{\downarrow,2r+1}) \nonumber &\\
    &= (\mu^z_{c,2r-1/2} \mu^z_{c,2r+3/2}) \cdot (\mu^z_{n,2r-1/2} \mu^z_{n,2r+3/2}) \cdot (\a^z_{\downarrow,2r} \a^z_{\downarrow,2r+1}) \nonumber &\\
    &= \tau'^x_{c,2r} \tau'^x_{c,2r+1} \tau'^x_{n,2r+1} \nonumber &\\
    &= \tau''^x_{c,2r} \tau''^x_{c,2r+1} \tau''^x_{n,2r+1} , &
     \end{flalign}
    \begin{flalign}
   \quad \s^x_{\uparrow,2r-1} \s^x_{\uparrow,2r} &= (\s^x_{\uparrow,2r-1} \s^x_{\downarrow,2r-1})(\s^x_{\uparrow,2r} \s^x_{\downarrow,2r}) (\s^x_{\downarrow,2r-1} \s^x_{\downarrow,2r}) \nonumber &\\
    &= \tau''^x_{c,2r-1} \tau''^x_{c,2r} \tau^x_{n,2r}.&
\end{flalign}
We summarize the spin-spin interactions in Table~\ref{tab:spinspindict} below.

\begin{table}[h]
\centering
\caption{Mappings between neighboring spin-spin interactions and partons coupled with gauge fields. $\tau_{c,r}$ and $\tau_{n,r}$ are chargons and spinons on site $r$, respectively. $\g_{2r+1/2}$ is the $\mathbb{Z}_2$ gauge field on the link $(2r+1/2)$.}
\begin{tabular}{ c|c|c } 
\hline \hline
Local spins & domain-walls & Partons \\
\hline
$\s^z_{\uparrow,2r} \s^z_{\uparrow,2r+1}$ & $\mu^x_{c,2r+1/2}$ & $\tau^z_{c,2r} \g^z_{2r+1/2} \tau^z_{c,2r+1}$ \\
$\s^z_{\uparrow,2r-1} \s^z_{\uparrow,2r}$ & $\mu^x_{c,2r-1/2} \mu^x_{n,2r-1/2}$ & $\tau^z_{n,2r-1} \g^z_{2r+1/2} \tau^z_{n,2r+1} \tau^z_{c,2r-1} \tau^z_{c,2r}$ \\
$\s^z_{\downarrow,2r} \s^z_{\downarrow,2r+1}$ & $\mu^x_{c,2r+1/2} \mu^x_{n,2r+1/2}$ & $\tau^z_{c,2r} \tau^z_{c,2r+1} \tau^z_{n,2r} \tau^z_{n,2r+2}$ \\
$\s^z_{\downarrow,2r-1} \s^z_{\downarrow,2r}$ & $\mu^x_{c,2r-1/2}$ & $\tau^z_{c,2r-1} \tau^z_{c,2r}$ \\
$\s^x_{\uparrow,r} \s^x_{\downarrow,r}$ & $\mu^z_{c,r-1/2} \a^z_{\uparrow,r} \a^z_{\downarrow,r} \mu^z_{c,r+1/2}$ & $\tau^x_{c,r}$ \\
$\s^x_{\uparrow,2r} \s^x_{\uparrow,2r+1}$ & $\mu^z_{n,2r-1/2} \a^z_{\uparrow,2r} \a^z_{\uparrow,2r+1} \mu^z_{n,2r+3/2}$ & $\tau^x_{n,2r+1}$ \\
$\s^x_{\uparrow,2r-1} \s^x_{\uparrow,2r}$ & $\mu^z_{c,2r-3/2} \mu^z_{n,2r-3/2} \a^z_{\uparrow,2r-1} \a^z_{\uparrow,2r} \mu^z_{c,2r+1/2} \mu^z_{n,2r+1/2}$ & $\tau^x_{c,2r-1} \tau^x_{c,2r} \tau^x_{n,2r}$ \\
$\s^x_{\downarrow,2r} \s^x_{\downarrow,2r+1}$ & $\mu^z_{c,2r-1/2} \mu^z_{n,2r-1/2} \a^z_{\downarrow,2r} \a^z_{\downarrow,2r+1} \mu^z_{c,2r+3/2} \mu^z_{n,2r+3/2}$ & $\tau^x_{c,2r} \tau^x_{c,2r+1} \tau^x_{n,2r+1}$ \\
$\s^x_{\downarrow,2r-1} \s^x_{\downarrow,2r}$ & $\mu^z_{n,2r-3/2} \a^z_{\downarrow,2r-1} \a^z_{\downarrow,2r} \mu^z_{n,2r+1/2}$ & $\tau^x_{n,2r}$ \\
\hline \hline
\end{tabular}
\label{tab:spinspindict}
\end{table}

The single spin representation in terms of partons in Table~\ref{tab:dictionary} can be read off easily from these spin-spin interactions.
\newpage
\end{widetext}

\section{Details of redefining composite symmetries and fluxes in Eq.~(\ref{eqn:redefine})}
\label{append:composite}
We derive Eq.~(\ref{eqn:redefine}) by redefining Charge and Spin symmetries as
\begin{align}
g'^\up_x &= g^\up_x g^\down_x, \ g'^\down_x = g^\down_x g_z, \ g'_z = g_z.
\end{align}
Table~\ref{tab:chargeredefine} below shows the quantum numbers of local spins under the redefined symmetries, leading to the minimal couplings
\begin{equation}
\begin{split}
\s^z_{\up,r} \s^z_{\up,r+1} &\rightarrow \s^z_{\up,r} \O'^z_{\text{C} \up, \tilde{r}} \s^z_{\up,r+1}, \\
\s^z_{\down,r} \s^z_{\down,r+1} &\rightarrow \s^z_{\down,r} \O'^z_{\text{C}\up,\tilde{r}} \O'^z_{\text{C} \down, \tilde{r}} \s^z_{\down,r+1}, \\
\s^x_{\up,r} \s^x_{\up,r+1}  &\rightarrow \s^x_{\up,r} \O'^z_{\text{S}, \tilde{r}} \O'^z_{\text{C}\down, \tilde{r}} \s^x_{\up,r+1}, \\
\s^x_{\down,r} \s^x_{\down,r+1} &\rightarrow \s^x_{\down,r} \O'^z_{\text{S}, \tilde{r}} \O'^z_{\text{C}\down,\tilde{r}} \s^x_{\down,r+1}.
\end{split}
\end{equation}
Consequently, the gauge fluxes before and after redefinition satisfy the relations 
\begin{align}
\Phi'_{\text{C}\up} &= \Phi_{\text{C}\up}, \  \Phi'_{\text{C}\down} = \Phi_{\text{C} \down} \Phi_{\text{C}\up}, \  \Phi'_{\text{S}} = \Phi_{\text{S}}\Phi_{\text{C} \up} \Phi_{\text{C}\down}.
\end{align}

\begin{table}[h]
\centering
\caption{Symmetry charges that are carried by local spin operators. A $\checkmark$ indicates that the operator is charged under this symmetry.}
\begingroup
\renewcommand{\arraystretch}{1}
\begin{tabular}{ c|ccc } 
\hline \hline
  & $g'^\up_x$ & $g'^\down_x$ & $g'_z$  \\
\hline
$\s^z_{\up,r}$ & \checkmark &  & \\
$\s^z_{\down,r}$ & \checkmark & \checkmark & \\
$\s^x_{\up,r}$ & & \checkmark & \checkmark\\
$\s^x_{\down,r}$ & & \checkmark &  \checkmark \\
\hline \hline
\end{tabular}
\endgroup
\label{tab:chargeredefine}
\end{table}

\section{Details of the SPT-II phase}
\label{appendix:cssspt}
To facilitate the analysis, we perform a rotation $\s^y_r \leftrightarrow \s^x_r$, followed by a Kramers-Wannier duality transformation, resulting in
\begin{equation}
\begin{split}
    &H_d = -J_{\text{C}} \sum_r \left( \tilde{\s}^x_{\uparrow,2 \tilde{r}} \O^z_{\text{C}\up,2\tilde{r}} + \tilde{\s}^x_{\downarrow,2\tilde{r}-1}  \O^z_{\text{C}\down,2\tilde{r}-1}\right) \\
    & - J_{\text{S}} \sum_r \left[ \tilde{\s}^z_{\downarrow,2 \tilde{r}-2} \tilde{\s}^x_{\uparrow,2\tilde{r}-1} (\O^z_{\text{C}\up,2\tilde{r}-1} \O^z_{\text{C}\down,2\tilde{r}-1}  \O^z_{\text{S},2\tilde{r}-1} ) \tilde{\s}^z_{\downarrow,2\tilde{r}} \right.  \\
    &~ \left. + \tilde{\s}^z_{\uparrow,2\tilde{r}-1} \tilde{\s}^x_{\downarrow,2\tilde{r}} (\O^z_{\text{C}\up,2\tilde{r}} \O^z_{\text{C}\down,2\tilde{r}}  \O^z_{\text{S},2\tilde{r}} ) \tilde{\s}^z_{\uparrow,2\tilde{r}+1} \right], \\
    &H'_d = -J'_{\text{C}} \sum_r \left( \tilde{\s}^x_{\uparrow,2 \tilde{r}-1} \O^z_{\text{C}\up,2\tilde{r}-1} + \tilde{\s}^x_{\downarrow,2\tilde{r}}  \O^z_{\text{C}\down,2\tilde{r}}\right) \\
    & - J'_{\text{S}} \sum_r \left[ \tilde{\s}^z_{\downarrow,2 \tilde{r}-1} \tilde{\s}^x_{\uparrow,2\tilde{r}} (\O^z_{\text{C}\up,2\tilde{r}} \O^z_{\text{C}\down,2\tilde{r}}  \O^z_{\text{S},2\tilde{r}} ) \tilde{\s}^z_{\downarrow,2\tilde{r}+1} \right.  \\
    &~ \left. + \tilde{\s}^z_{\uparrow,2\tilde{r}} \tilde{\s}^x_{\downarrow,2\tilde{r}+1} (\O^z_{\text{C}\up,2\tilde{r}+1} \O^z_{\text{C}\down,2\tilde{r}+1}  \O^z_{\text{S},2\tilde{r}+1} ) \tilde{\s}^z_{\uparrow,2\tilde{r}+2} \right].
\end{split}
\end{equation}
Notice that the $(J_{\text{C}},J'_{\text{S}})$ terms and the $(J_{\text{S}}, J'_{\text{C}})$ terms form two decoupled chains. We denote
\begin{equation}
\begin{split}
    H^{XY}_A &= \text{$(J_{\text{C}}, J'_{\text{S}})$ terms}, \quad   H^{XY}_B = \text{$(J'_{\text{C}}, J_{\text{S}})$ terms}.
\end{split}
\end{equation}
and define
\begin{equation}
\begin{split}
    \tilde{\m{X}}^A_{2\tilde{r}} &\equiv \tilde{\s}^x_{\uparrow,2\tilde{r}}, \qquad \tilde{\m{X}}^A_{2\tilde{r}-1} \equiv \tilde{\s}^x_{\downarrow,2\tilde{r}-1}, \\
    \tilde{\m{X}}^B_{2\tilde{r}} &\equiv \tilde{\s}^x_{\downarrow,2\tilde{r}}, \qquad \tilde{\m{X}}^B_{2\tilde{r}-1} \equiv \tilde{\s}^x_{\uparrow,2\tilde{r}-1}, \\
    \tilde{\m{Z}}^A_{2\tilde{r}} &\equiv \tilde{\s}^z_{\uparrow,2\tilde{r}}, \qquad \tilde{\m{Z}}^A_{2\tilde{r}-1} \equiv \tilde{\s}^z_{\downarrow,2\tilde{r}-1}, \\
    \tilde{\m{Z}}^B_{2\tilde{r}} &\equiv \tilde{\s}^z_{\downarrow,2\tilde{r}}, \qquad \tilde{\m{Z}}^B_{2\tilde{r}-1} \equiv \tilde{\s}^z_{\uparrow,2\tilde{r}-1},
\end{split} \label{eqn:relabeling}
\end{equation}
which leads to
\begin{equation}
\begin{split}
    H^{XY}_A &= -J_{\text{C}} \sum_{r \text{ even}} (\tilde{\m{X}}^A_{\tilde{r}} \O^z_{\text{C}\up,\tilde{r}} + \tilde{\m{X}}^A_{\tilde{r}-1} \O^z_{\text{C}\down,\tilde{r}-1} ) \\
&~ - J'_{\text{S}} \sum_{r} \tilde{\m{Z}}^A_{\tilde{r}-1} \tilde{\m{X}}^A_{\tilde{r}} (\O^z_{\text{C}\up,\tilde{r}} \O^z_{\text{C}\down,\tilde{r}}  \O^z_{\text{S},\tilde{r}} ) \tilde{\m{Z}}^A_{\tilde{r}+1}, \\
    H^{XY}_B &= -J'_{\text{C}} \sum_{r \text{ even}}( \tilde{\m{X}}^B_{\tilde{r}-1} \O^z_{\text{C}\up, \tilde{r}-1} + \tilde{\m{X}}^B_{\tilde{r}} \O^z_{\text{C}\down,\tilde{r}}) \\
&~ - J_{\text{S}} \sum_{r} \tilde{\m{Z}}^B_{\tilde{r}-1} \tilde{\m{X}}^B_{\tilde{r}} (\O^z_{\text{C}\up,\tilde{r}} \O^z_{\text{C}\down,\tilde{r}}  \O^z_{\text{S},\tilde{r}} ) \tilde{\m{Z}}^B_{\tilde{r}+1}.
\end{split}
\end{equation}
A second Kramers-Wannier duality and subsequently a basis rotation $\m{Y}_r \leftrightarrow (-1)^r \m{X}_r$ lead to
\begin{equation}
\begin{split}
    H^{XY}_A &= -J_{\text{C}} \sum_{r \text{ even}} (\m{Z}^A_r \O^z_{\text{C}\up,\tilde{r}} \m{Z}^A_{r+1} + \m{Z}^A_{r-1} \O^z_{\text{C}\down,\tilde{r}-1} \m{Z}^A_{r}) \\
&~ - J'_{\text{S}} \sum_r \m{X}^A_r (\O^z_{\text{C}\up,\tilde{r}} \O^z_{\text{C}\down,\tilde{r}}  \O^z_{\text{S},\tilde{r}} ) \m{X}^A_{r+1}, \\
    H^{XY}_B &= -J'_{\text{C}} \sum_{r \text{ even}} (\m{Z}^B_{r-1} \O^z_{\text{C}\up,\tilde{r}-1} \m{Z}^B_{r} + \m{Z}^B_{r} \O^z_{\text{C}\down,\tilde{r}} \m{Z}^B_{r+1} ) \\
&~ - J_{\text{S}} \sum_r \m{X}^B_r (\O^z_{\text{C}\up,\tilde{r}} \O^z_{\text{C}\down,\tilde{r}}  \O^z_{\text{S},\tilde{r}} ) \m{X}^B_{r+1},
\end{split}
\end{equation}
which describe two decoupled XY models. The interaction term in Eq.~(\ref{eqn:intVBS}) becomes
\begin{equation}
\begin{split}
    H_{\text{int}} &= \sum_r (\s^z_{\up,r} \s^z_{\up,r+1} - \s^z_{\up,r} \s^z_{\up,r+1} \s^y_{\down,r} \s^y_{\down,r} ) \\
    &\ \times (\s^z_{\down,r} \s^z_{\down,r+1} - \s^y_{\up,r} \s^y_{\up,r+1} \s^z_{\down,r} \s^z_{\down,r+1}) \label{eqn:localHint}
    \end{split}
\end{equation}
where we used the relations between local spins and their dual spins
\begin{equation}
\begin{split}
    &\m{Z}^A_r \m{Z}^A_{r+1} = \s^z_{\uparrow,2r} \s^z_{\uparrow,2r+1} + \s^z_{\downarrow,2r-1} \s^z_{\downarrow,2r}, \\
    &\m{Z}^B_r \m{Z}^B_{r+1} = \s^z_{\downarrow,2r} \s^z_{\downarrow,2r+1} + \s^z_{\uparrow,2r-1} \s^z_{\uparrow,2r}, \\
    &\m{X}^A_r \m{X}^A_{r+1} \rightarrow \m{Y}^A_r \m{Y}^A_{r+1} \\
    &= \s^z_{\uparrow,2r} \s^z_{\uparrow,2r+1} \s^y_{\downarrow,2r} \s^y_{\downarrow,2r+1} + \s^z_{\downarrow,2r-1} \s^z_{\downarrow,2r} \s^y_{\uparrow,2r-1} \s^y_{\uparrow,2r} , \\
    &\m{X}^B_r \m{X}^B_{r+1} \rightarrow \m{Y}^B_r \m{Y}^B_{r+1}  \\
    &=  \s^z_{\uparrow,2r-1} \s^z_{\uparrow,2r} \s^y_{\downarrow,2r-1} \s^y_{\downarrow,2r} + \s^z_{\downarrow,2r} \s^z_{\downarrow,2r+1} \s^y_{\uparrow,2r} \s^y_{\uparrow,2r+1}.
\end{split}
\end{equation}
Notice that $\m{Z}^A_r \m{Z}^A_{r+1} \m{X}^B_r \m{X}^B_{r+1}$ is the product of $(J_\text{C},J_\text{S})$ terms, while $\m{X}^A_r \m{X}^A_{r+1} \m{Z}^B_r \m{Z}^B_{r+1}$ is the product of $(J'_\text{C},J'_\text{S})$ terms. Notice also that $(J'_c,J'_s)$ terms are the translational partners of $(J_\text{C},J_\text{S})$ terms. Consequently, the sum of the two terms is translationally invariant. 

\section{Details of phase transitions between SPT-II and $y$FM/$z$FM-I phases in Section~\ref{subsec:Txenhanced}}
\label{appendix:transition}

\subsubsection{Transition between yFM and SPT-II}
Transition 3 in Fig.~\ref{fig:translation} connects the $y$FM phase (chargon insulator, spinon SPT state) and the SPT-II phase (chargon condensate, spinon SPT state). Both phases are accessible by perturbing the gapless parent Hamiltonian in the dual spin representation introduced in Eq.~(\ref{eqn:dualbasis})
\begin{align}
    \tilde{H}_\text{p} &= - \sum_{r,a = A,B}  ( \m{Z}^a_r \m{Z}^a_{r+1} + \m{X}^a_r \m{X}^a_{r+1} - \tilde{\D} \m{Y}^a_r \m{Y}^a_{r+1} ).
\end{align}
Specifically, $y$FM arises upon perturbing $\tilde{H}_\text{p}$ with $\m{X}^A \m{X}^B$ and SPT-II as prescribed in Section~\ref{sec:partontranslation}. As discussed for Transition 1, the lattice model $\tilde{H}_\text{p}$ corresponds to the Hamiltonian density
\begin{equation}
\tilde{{\cal H}}_{\text{LL}} = \frac{\tilde{v}_F}{2\pi}\sum_{a=A,B}[\tilde{K}_a^{-1}(\partial_x \Phi_a)^2+\tilde{K}_a(\partial_x \Theta_a)^2], \label{eqn:LL2}
\end{equation}
The long-wavelength variables $\Phi,\Theta$ satisfy the canonical commutation relations
\begin{align}
    \left[ \p_x \Theta_a(x), \Phi_{a'}(x') \right] &= i\pi \d_{aa'} \d(x-x'), \quad a,a'=A,B,
\end{align}
and encode dual spin operators via
\begin{equation}
\begin{split}
    \m{X}^a_r + i\m{Z}^a_r &\sim e^{-i \Phi_a} \left[ \m{C}_1 + \m{C}_2 (-1)^r \cos{2 \Theta_a} \right], \\
    \m{Y}^a_r &\sim \frac{2\p_x \Theta_a}{\pi} + \m{C}_3 (-1)^r \sin{2\T_a}, \ a = A,B~,
\end{split}
\end{equation}
where $\m{C}_{1,2,3}$ are normalization factors. 

In this dual representation, a sine-Gordon term that leads to the $y$FM phase is
\begin{equation}
\begin{split}
\d \m{H}_{y\text{FM}} %&= -J_{\text{C}} \sum_r \m{X}^A_r \m{X}^B_r - J_\text{S} \sum_r (\m{X}^A_r \m{X}^A_{r+1} + \m{X}^B_r \m{X}^B_{r+1}) \\
    &=- 2 J_y \cos{\Phi_A} \cos{\Phi_B}. %- J_\text{S} \left( \cos{2\varphi_A} + \cos{2\varphi_B} \right),
    \end{split}
\end{equation}
whose ground states exhibit $\langle \cos{\Phi_A} \rangle = \langle \cos{\Phi_B} \rangle \neq 0$, breaking $g^\up_x, g^\down_x, g_z$ individual while preserving any product of two of them. 
The SPT-II phase arises from the pruturbation
\begin{align}
    \d \m{H}_{\text{SPT-II}} &= + 2J_{\text{SPT-II}} \cos{(2\T_A)} \cos{(2\T_B)}, \label{eqn:sptii}
\end{align}
which realizes a phase with $\langle \cos{(2\T_A)} \rangle = - \langle \cos{(2\T_B)} \rangle \neq 0$, breaking translation symmetry. 

As discussed in Section~\ref{subsec:tt}, to understand the transition between $y$FM and SPT-II phases, we perform the transformation $\Phi_\pm \equiv \frac{1}{2}(\Phi_A \pm \Phi_B)$. The transition is thus described by
\begin{equation}
\begin{split}
     \m{H}^{(3)}_{\text{transition}} &= \tilde{H}_{\text{LL}} + 
    \d \m{H}_{y \text{FM}} + \d \m{H}_{\text{SPT-II}} \\
    &= \tilde{H}_{\text{LL}} + \sum_{\kappa = \pm} (-J_y \cos{2\Phi_\kappa} + J_{\text{SPT-II}} \cos{2\T_\kappa} ). \label{eqn:transitionCISSPT}
    \end{split}
\end{equation}
The second line describes independent Ising transitions for $\kappa = \pm$. Following similar analyses from Eq.~(\ref{eqn.xfmspt1}) to (\ref{eqn:Hint1Majorana}), we conclude that the critical points of $\m{H}^{(3)}_{\text{transition}}$ 
at $J_{\text{SPT-II}} = (\pm) J_y$ describe Ising $\times$ Ising transitions with central charge $c=1$ characterized by two massless Majorana fermions. Marginal interactions move the critical point away from the Ising $\times$ Ising point on the AT line. Explicitly breaking translation invariance reduces the critical point to a single Ising transition with central charge $c=\frac{1}{2}$. 
\newline

\subsubsection{Transition between zFM-I and SPT-II}
Transition 4 in Fig.~(\ref{fig:translation}) connects the $z$FM-I phase (chargon condensate, spinon condensate) to the SPT-II phase (chargon condensate, spinon SPT state), and mirrors Transition 2 in the dual representation. Specifically, the transition between $z$FM-I and SPT-II phases is captured by
\begin{align}
    \m{H}^{(4)}_{\text{transition}} &= \tilde{\m{H}}_{\text{LL}} + \d \m{H}_{z\text{FM-I}} + \d \m{H}_{\text{SPT-II}} 
\end{align}
with $\d \m{H}_{\text{SPT-II}}$ given in Eq.~(\ref{eqn:sptii}) and  
\begin{equation}
    \begin{split}
    \d \m{H}_{z\text{FM-I}}
     &=- J_z \left[ \sin{(2\Phi_A)} + \sin{(2\Phi_B)} \right].  
    \end{split}
\end{equation}
\begin{equation}
\d \m{H}_{\text{SPT-II}} = +2 J_{\text{SPT-II}} \cos{(2\T_A)} \cos{(2\T_B)}.
\end{equation} 
Following similar analyses for Transition 2 from below Eq.~(\ref{eqn:zfmi}), we expect that the critical point between $z$FM-I and SPT-II has a central charge $c=3/2$.

%\bibliography{parton}

\end{document}